\documentclass[twocolumn,superscriptaddress,showpacs,preprintnumbers,amsmath,amssymb]{revtex4}

\usepackage[dvipdfm]{graphicx}
\usepackage{dcolumn}
\usepackage{bm}
\usepackage{longtable}

\usepackage[dvips]{color} 

{%
   \begin{list}{\parbox{1zw}{$\bullet$}}
   {%
      \setlength{\itemsep}{0em}
      \setlength{\parsep}{0em}
   }
}{%
   \end{list}%
}

\begin{document}

\preprint{APS/123-QED}

\title{Analyzing power in elastic scattering of $^6$He from polarized
proton target at 71~MeV/nucleon}

\author{S.~Sakaguchi}
 \altaffiliation{Present address: Department of Physics, Kyushu
 University, Fukuoka 812-8581, Japan;\\ Electronic address:
 {\tt sakaguchi@phys.kyushu-u.ac.jp}}
\affiliation{Center for Nuclear Study, University of Tokyo,
 Tokyo 113-0001, Japan}
\author{Y.~Iseri}
\affiliation{Chiba-Keizai College, Chiba 263-0021, Japan}
\author{T.~Uesaka}
\affiliation{Center for Nuclear Study, University of Tokyo, 
 Tokyo 113-0001, Japan}
\author{M.~Tanifuji}
\affiliation{Science Research Center, Hosei University, Tokyo 102-8160, Japan}
\author{K.~Amos}
\affiliation{School of Physics, University of Melbourne, Melbourne, Australia}
\author{N.~Aoi}
\affiliation{RIKEN Nishina Center, Saitama 351-0198, Japan}
\author{Y.~Hashimoto}
\affiliation{Department of Physics, Tokyo Institute of Technology, Tokyo 152-8551, Japan}
\author{E.~Hiyama}
\affiliation{RIKEN Nishina Center, Saitama 351-0198, Japan}
\author{M.~Ichikawa}
\affiliation{Cyclotron \& Radioisotope Center, Tohoku University, Miyagi 980-8578, Japan}
\author{Y.~Ichikawa}
\affiliation{Department of Physics, University of Tokyo, Tokyo 113-0033, Japan}
\author{S.~Ishikawa}
\affiliation{Science Research Center, Hosei University, Tokyo 102-8160, Japan}
\author{K.~Itoh}
\affiliation{Department of Physics, Saitama University, Saitama 338-8570, Japan}
\author{M.~Itoh}
\affiliation{Cyclotron \& Radioisotope Center, Tohoku University, Miyagi 980-8578, Japan}
\author{H.~Iwasaki}
\affiliation{Department of Physics, University of Tokyo, Tokyo 113-0033, Japan}
\author{S.~Karataglidis}
\affiliation{Department of Physics, University of Johannesburg,
P.O. box 924, Auckland Park, 2006 South Africa}                            
\author{T.~Kawabata}
\affiliation{Center for Nuclear Study, University of Tokyo, 
 Tokyo 113-0001, Japan}
\author{T.~Kawahara}
\affiliation{Department of Physics, Toho University, Chiba, Japan}
\author{H.~Kuboki}
\affiliation{Department of Physics, University of Tokyo, Tokyo 113-0033, Japan}
\author{Y.~Maeda}
\affiliation{Center for Nuclear Study, University of Tokyo, 
 Tokyo 113-0001, Japan}
\author{R.~Matsuo}
\affiliation{Cyclotron \& Radioisotope Center, Tohoku University, Miyagi 980-8578, Japan}
\author{T.~Nakao}
\affiliation{Department of Physics, University of Tokyo, Tokyo 113-0033, Japan}
\author{H.~Okamura}
\altaffiliation{Deceased.}
\affiliation{Research Center for Nuclear Physics, Osaka University, Osaka 567-0047, Japan}
\author{H.~Sakai}
\affiliation{Department of Physics, University of Tokyo, Tokyo 113-0033, Japan}
\author{Y.~Sasamoto}
\affiliation{Center for Nuclear Study, University of Tokyo, 
 Tokyo 113-0001, Japan}
\author{M.~Sasano}
\affiliation{Department of Physics, University of Tokyo, Tokyo 113-0033, Japan}
\author{Y.~Satou}
\affiliation{Department of Physics, Tokyo Institute of Technology, Tokyo 152-8551, Japan}
\author{K.~Sekiguchi}
\affiliation{RIKEN Nishina Center, Saitama 351-0198, Japan}
\author{M.~Shinohara}
\affiliation{Department of Physics, Tokyo Institute of Technology, Tokyo 152-8551, Japan}
\author{K.~Suda}
\affiliation{Center for Nuclear Study, University of Tokyo, 
 Tokyo 113-0001, Japan}
\author{D.~Suzuki}
\affiliation{Department of Physics, University of Tokyo, Tokyo 113-0033, Japan}
\author{Y.~Takahashi}
\affiliation{Department of Physics, University of Tokyo, Tokyo 113-0033, Japan}
\author{A.~Tamii}
\affiliation{Research Center for Nuclear Physics, Osaka University, Osaka 567-0047, Japan}
\author{T.~Wakui}
\affiliation{Cyclotron \& Radioisotope Center, Tohoku University, Miyagi 980-8578, Japan}
\author{K.~Yako}
\affiliation{Department of Physics, University of Tokyo, Tokyo 113-0033, Japan}
\author{M.~Yamaguchi}
\affiliation{Graduate School of Medicine, Gunma University, Gunma 229-8510, Japan}
\author{Y.~Yamamoto}
\affiliation{Tsuru University, Yamanashi 402-8555, Japan}
\date{\today}

\begin{abstract}
The vector analyzing power has been measured for the elastic
scattering of neutron-rich $^6$He from polarized protons at 71~MeV/nucleon
making use of a
newly constructed solid polarized proton target operated in a low
magnetic field and at high temperature. 
Two approaches based on local one-body potentials were
applied to investigate the spin-orbit interaction between
a proton and a $^6$He nucleus.
An optical model analysis revealed that the spin-orbit
potential for $^6$He is characterized by a shallow and long-ranged
shape compared with the global systematics of stable
 nuclei.
A semi-microscopic analysis with a $\alpha$+$n$+$n$ cluster folding model 
suggests that 
the interaction between a proton and the $\alpha$ core 
is essentially important 
in describing the 
$p+^6$He elastic 
scattering.
The data are also compared with fully microscopic analyses 
using non-local optical potentials based on nucleon-nucleon $g$-matrices.
\end{abstract}


\maketitle

\section{Introduction}

Spin-orbit coupling in atomic nuclei is an essential feature in
understanding any reaction and nuclear structure related to it.
One of the direct manifestations of that spin-orbit coupling in nuclear
reactions, is the polarization phenomenon in nucleon elastic
scattering~\cite{Oxley53,Chamberlain56,Fermi54}.
Characteristics of the spin-orbit coupling between a nucleon and stable
nuclei have been well established by analyses of measured vector analyzing 
powers in the elastic scattering of polarized nucleons on various targets
over a wide range of incident energies~\cite{Varner91,Koning03,Brieva78,Amos00}.

On the other hand, the spin-orbit coupling of a nucleon with unstable
nuclei might be considerably different from that with the stable nuclei.  
Some neutron-rich nuclei with small binding energies are known to
have very extended neutron distributions~\cite{Tanihata85}.
Since the spin-orbit coupling is essentially a surface effect, it is
natural to  expect that 
the diffused density distribution of a neutron-rich nucleus may 
significantly effect the radial shape and depth of the spin-orbit potential.
The purpose of this work is to investigate the 
characteristics of the spin-orbit potential between a proton and 
$^6$He; a typical neutron-rich nucleus.

Experimental determination of the spin-orbit potential strongly owes to
measurements and analyses of the vector analyzing powers.
However, until recently, analyzing power data were not
obtained in the scattering which involves unstable nuclei.
This was mainly due to the lack of a polarized proton target that is
applicable to radioactive ion (RI) beam experiments.
RI-beam experiments induced by light ions are usually carried out under
inverse-kinematics conditions, where energies of recoil protons can be as
low as 10~MeV.
Conventional polarized proton targets~\cite{Crabb97,Goertz02}, based on 
the dynamic nuclear polarization method, 
require a high magnetic field and low temperature 
such as 2.5~T and 0.5~K, respectively.
It is impossible to detect the low-energy 
recoil protons with sufficient angular resolution under these extreme
conditions.
For the application in RI-beam experiments,
we have constructed a solid polarized proton target which
can be operated under low magnetic field of 0.1~T and at high
temperature of 100~K~\cite{Wakui04,Uesaka04,Wakui05,Hatano05,WakuiPST}.
The electron polarization in photo-excited aromatic
molecules is used to polarize the protons~\cite{Sloop81,Henstra88}.
A high proton polarization of about 20\% can be achieved in relatively
``relaxed'' operating conditions described above, since the magnitude of
the electron
polarization is almost independent of the magnetic field strength and
temperature.

We have measured the vector analyzing power for the $p+^6$He elastic
scattering at 71~MeV/nucleon~\cite{Uesaka10} using the solid polarized
proton target, newly constructed for RI-beam experiments.
$^6$He is suitable for the present study since
it has a spatially extended distribution due to a small binding energy.
In addition, from an experimental viewpoint, 
the $p+^6$He elastic scattering measurement is relatively easy to perform
since $^6$He does not have a bound excited state.
This allows us to identify the elastic-scattering event only by
detecting $^6$He and a proton in coincidence.
%
%
%
The analyzing powers thus measured are 
the first data set that can be used for quantitative evaluation of
the spin-orbit interaction between a proton and an unstable $^6$He
nucleus. 
The essence of these measurements has been published in
Ref.~\cite{Uesaka10} together with two kinds of theoretical analyses by
folding models; one assumes a fully antisymmetrized large-basis shell
model for $^6$He with the $g$-matrix interaction and the other an
$\alpha$+$n$+$n$ cluster model for $^6$He with a $p$-$n$ effective interaction
and a realistic $p$-$\alpha$ static potential.
The main purpose of the present paper is to give more details of the 
experiment and present an additional analysis of the experimental data 
using  a one-body
$p$-$^6$He optical potential.
The analysis exhibits remarkable characteristics for the spin-orbit part
of that potential.
Then it becomes important to investigate if such a potential can be
derived theoretically from any model of $^6$He.
As the first approach we examined the $\alpha$+$n$+$n$ folding potential in
more detail, since important contributions of the $\alpha$ cluster are
suggested by the fact that 
the measured $A_y$ for $^6$He is similar to that for
$^4$He~\cite{Uesaka10}, when plotted versus the momentum transfer of the
scattering.
To identify effects of the clusterization, we also calculated the
$p$-$^6$He folding potential for a $2p$+$4n$ non-cluster model of $^6$He and
compared the results with those of the $\alpha$+$n$+$n$ cluster model.
Hereafter, they are referred to as the $\alpha nn$ cluster folding (CF)
model and nucleon folding (NF) one, respectively.
In addition, the data are also compared with fully microscopic
calculations using non-local optical potentials.
In this model, non-locality of the $p$-$^6$He interaction,
a consequence of the Pauli principle leading to  
nucleon exchange scattering amplitudes,
is taken into account explicitly.
Three sets of single-particle wave functions, as well as
the required one-body density matrix elements determined 
from a large-basis
shell model for $^6$He, have been used in these calculations.

The present paper is subdivided as follows.
In Section 2, details of the experimental method are described.
In Section 3, the method of the data reduction is presented.
Section 4 deals with the phenomenological optical model analysis.
Section 5 is devoted to the details of the $\alpha n n$ cluster folding
calculation and the 
nucleon 
folding calculation.
In Section 6, the data are compared with the analysis by the non-local
$g$-folding optical potentials.
Finally, a short summary of the obtained results is given in Section 7.

\section{Experiment}

\subsection{Experimental setup}

The experiment was carried out at the RIKEN Accelerator
Research Facility (RARF).
The $^6$He beam was produced
through the projectile fragmentation of a $^{12}$C beam with
an energy of 92~MeV/nucleon bombarding a primary target.
As that primary, we used a rotating $^9$Be target~\cite{Yoshida08}
to avoid heat damage by the beam. 
A thickness of the target was 1480 mg/cm$^2$.
The $^6$He particles were separated by the RIKEN Projectile-fragment Separator
(RIPS)~\cite{Kubo92} based on the magnetic rigidity and the energy loss of 
fragments.
The energy of the $^6$He beam was 70.6$\pm$1.4~MeV/nucleon at the center of
the secondary target.
The purity of the beam was 95\%.

\begin{figure*}[htbp]
 \begin{center}
  \includegraphics[width=0.69\linewidth]{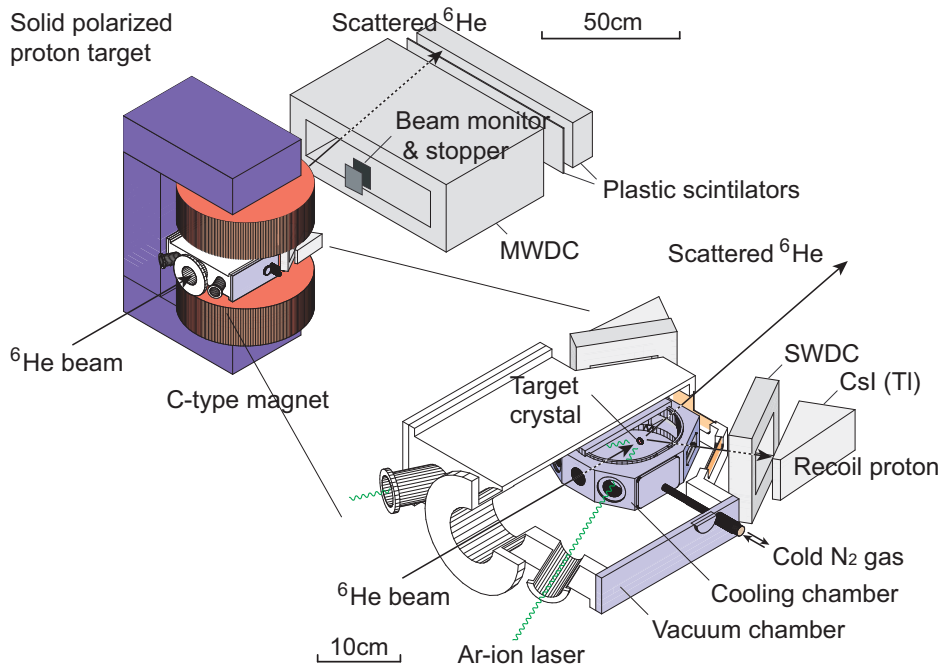}
  \caption{(Color online) Experimental setup of the secondary target and detectors is
  shown.}
  \label{fig:setup}
 \end{center}
\end{figure*}

The solid polarized proton target was placed at the final focal plane of
RIPS.
Figure~\ref{fig:setup} illustrates the experimental setup of the target
and detectors.
The most prominent advantage of the target is its relaxed operating conditions,
i.e. a low magnetic field of 0.1~T and high temperature of 100~K.
These conditions allow us to detect recoil protons 
whose energies are as low as 10~MeV.
Details of the target will be described in the following subsection.

A detector system consisted of two subsystems: one for scattered particles and
the other for recoil protons.
Detection of the recoil protons with energies as low as 10~MeV is
essential for the selection of the elastic-scattering events.
The scattering angles of protons were determined by single-wire drift
chambers (SWDC).
The SWDCs were placed 138.5~mm away from the target on both left and
right sides of the beam axis as shown in Fig.~\ref{fig:setup}.
They covered an angular region of 39$^{\circ}$--71$^{\circ}$ (horizontal) and
$\pm9.7^{\circ}$ (vertical) in the laboratory system.
Their position resolution and detection efficiency were found to be 
2.6~mm (FWHM) and 99.3\%.
For the measurement of the total energy of protons, we used CsI(Tl)
scintillation detectors.
They were placed just behind the SWDCs.
Light output from the CsI(Tl) crystal was detected by photo-multiplier
tubes.
The front side of the CsI(Tl) scintillator was covered by the thin
carbon-aramid film with a thickness of 12~$\mu$m.
Material thickness of the film, the SWDC, and air between the detectors was
24~mg/cm$^2$ in total.
Energy loss of 10~MeV protons in these materials is 1.2~MeV, which does
not prevent the detection.

A multi-wire drift chamber (MWDC) was used to 
reconstruct the trajectories of scattered particles.
Scattering position on the secondary target was determined from the
reconstructed trajectory.
The MWDC was placed at 880~mm downstream of the target.
It has a sensitive area of 640~mm (horizontal) $\times$ 160~mm
(vertical) and covered an angular region of $\pm$16$^{\circ} \times
\pm$4$^{\circ}$ in the laboratory system.
The configuration of the planes of the MWDC is X-Y-X'-Y'-X'-Y'-X-Y,
where ``X(Y)-plane'' has anode-wires oriented along the 
vertical (horizontal) axis.
The planes with primes are displaced with respect to the ``unprimed''
planes by half the cell size.
The cell size is 20~mm$\times$20~mm for the X-plane and 
10~mm$\times$10~mm for the Y-plane.
The material of the anode wire is gold-plated tungsten with a diameter
of 30~$\mu$m.
Negative high voltages were applied to the cathode and potential wires:
$-$2.85~kV for the X (X')-planes and $-$2.15~kV for the Y (Y')-planes.
A gas mixture of Ar (50\%) and C$_2$H$_6$ (50\%) was used.
Position resolution and detection efficiency of the MWDC were found to
be 0.2~mm (FWHM) and 99.8\%.
For identification of scattered particles, we used a plastic
scintillation detector array placed just behind the MWDC.
The first and second layers with thicknesses of 5~mm and 100~mm provided
information of the energy loss and the total energy of scattered particles.
The total number of the beam particles was counted with a beam monitor 
placed between the secondary target and the MWDC.
A 50~mm$^H\times$50~mm$^V\times$10~mm$^D$ plastic scintillator was used
for the beam monitor.
A beam stopper made of a copper block was placed just behind the beam
monitor.

\subsection{Solid polarized proton target}
The solid polarized proton target, used in the measurement, can be operated
in a low magnetic field of 0.1~T and at a high temperature of
100~K.
These relaxed operation conditions allow us to detect low-energy
recoil protons without losing angular resolution.
This capability is indispensable to apply the target to scattering
experiments carried out under the inverse kinematics condition.
The proton polarization of about 20\% has been achieved~\cite{WakuiPST} under such
relaxed conditions by 
introducing a new polarizing method using electron polarization in
triplet states of photo-excited aromatic molecules~\cite{Sloop81,Henstra88}.
A single crystal of naphthalene (C$_{10}$H$_{8}$)
doped with a small amount of pentacene (C$_{22}$H$_{14}$) is used as the
target material.
Protons in the crystal are polarized by repeating a two-step process:
production of electron polarization and polarization transfer.
In the first step, pentacene molecules are optically excited to higher
singlet states. 
A small fraction of them decays to the first triplet state 
via the first excited singlet state by the so-called intersystem
crossing.
Here, electron population difference is spontaneously produced among
Zeeman sublevels of the triplet state~\cite{Sloop81}.
In the second step, the electron population difference between two
Zeeman sublevels, namely electron polarization, is transferred to
the proton polarization by the
cross-relaxation technique~\cite{Henstra88}.

As the target material, we used a single crystal of naphthalene doped
with 0.005 mol\% pentacene molecules.
The crystal was shaped into a thin disk whose diameter and thickness 
are 14 mm and 1 mm (116 mg/cm$^2$), respectively.
The number of hydrogens per unit area was 4.29$\pm$0.13$\times 10^{21}$/cm$^2$.
In order to reduce the relaxation rate, the target crystal was cooled
down to 100~K in a cooling chamber with the flow of cold nitrogen gas.
The cooling chamber was installed in another chamber as shown in
Fig.~\ref{fig:setup}.
Heat influx to the cooling chamber was reduced by the vacuum kept in the 
intervening space between these two chambers.
Each chamber has one window (6~$\mu$m-thick Havar foil) on the upstream
side for the incoming RI-beam, two glass windows for the laser irradiation, and
three windows (20~$\mu$m-thick Kapton
foil) on the left, right, and downstream sides for the detection of
recoil and scattered particles.

A static magnetic field was applied on the target crystal by a C-type 
electromagnet
to define the polarizing axis.
The gap and the diameter of the poles were 100~mm and 220~mm, respectively.
The strength of the magnetic field in the present experiment
was 91~mT; a value much higher than
that of the crystal field ($\approx$ 2~mT).
While the effects of the magnetic field on the scattering angles of 
$^6$He particles and protons were sufficiently small 
(about 0.07$^{\circ}$ and 0.2--0.8$^{\circ}$, respectively),
they were properly corrected in the data analysis.

The target crystal was irradiated by the light of two Ar-ion lasers with
a power of 25~W each in the multi-line mode.
Wavelengths of main components of the light were 514.5~nm (10~W) 
and 488.5~nm (8~W).
The laser light was pulsed by a rotating optical chopper.
Typically the pulse width and repetition rate were 12--14~$\mu$s and 1~kHz.
Microwave (MW) irradiation and a magnetic field sweep are required in the 
cross-relaxation method.
For the MW irradiation, 
the target crystal was installed in a resonator.
In order to detect low-energy recoil protons, we employed a thin
cylindrical loop-gap resonator (LGR~\cite{Ghim96})
made of 25 $\mu$m-thick Teflon film.
Copper stripes with a thickness of 4.4~$\mu$m were printed on both sides
of the film.
The MW frequency was 3.40~GHz.
The LGR was surrounded by a cylindrical MW shield made of
12~$\mu$m-thick aluminum foil.
For the cross-relaxation, the magnetic field was swept from 88~mT to
94~mT at the rate of 0.36~mT/$\mu$s, simultaneously with the MW
irradiation, by
applying a current to a small coil placed in the vicinity of the target
material.

Proton polarization was monitored 
during the experiment by the pulse NMR method.
A radio-frequency (RF) pulse with a frequency and a duration of
3.99~MHz and
2.2~$\mu$s was applied to a 19~mm$\phi$ NMR coil covering 
the target crystal.
The free induction decay (FID) signal was detected by the same coil.
We carried out the absolute calibration to relate the FID signal to the
proton polarization by measuring the
spin-asymmetry in the $p+^4$He elastic scattering.
Details of the calibration procedure are described in Appendix.~\ref{sec:calib}.

Devices located near to the target, namely
the LGR, the MW shield, the field sweeping coil, 
and the NMR coil, were fabricated with
hydrogen-free materials to prevent production of background
events.
Table~\ref{table:thick} shows the material thicknesses of the 
devices that recoil protons penetrate.
Energy losses of the 20~MeV protons in these materials are sufficiently
small for the detection as summarized in Table~\ref{table:thick}.

\begin{table*}[htbp]
\begin{center}
\begin{tabular}{c|cc}
\hline
\hline
Material & \hspace{3mm} Thickness (mg/cm$^2$) \hspace{3mm} &
 \hspace{3mm} Energy loss (MeV) \hspace{3mm}\\
\hline
\hspace{3mm} Target crystal (Naphthalene) \hspace{3mm} & 0 -- 336 & 0 -- 9.5\\
LGR (Teflon, Cu foil) & 9.3 & 0.2 -- 0.4\\
Microwave shield (Al foil) & 3.2 & 0.05 -- 0.1\\
Cooling gas (N$_2$) & 13.5 & 0.3 -- 0.6\\
Window (Kapton film) & 20 & 0.5 -- 1.0\\
\hline
Total & 46 -- 382 & 1.1 -- 11.6\\
\hline
\hline
\end{tabular}
\end{center}
\caption{
Thicknesses of the materials of target devices and energy losses
 of 20~MeV recoil protons in them.
}
\label{table:thick}
\end{table*}

\begin{figure}[htbp]
 \begin{center}
  \includegraphics[width=0.95\linewidth]{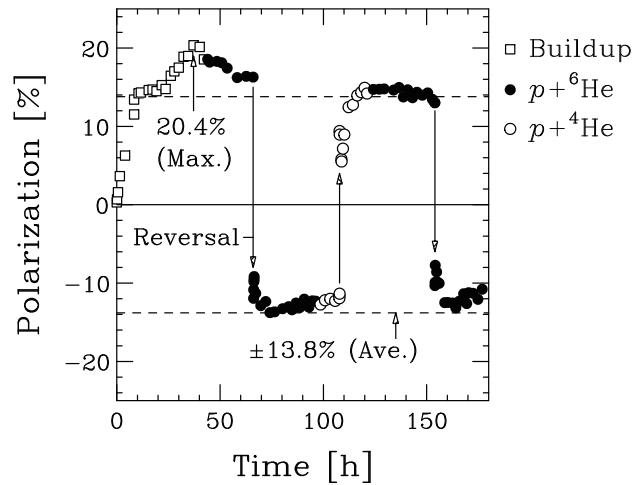}
  \caption{Target polarization is shown as a 
  function of time. The open squares, closed circles, and open circles
  indicate the target polarization during the polarization buildup, the
  $p+^6$He elastic-scattering measurement, and the $p+^4$He
  elastic-scattering measurement, respectively.}
  \label{fig:pol}
 \end{center}
\end{figure}

The target polarization during the experiment is shown in
Fig.~\ref{fig:pol} as a function of time.
The polarization was built up for the first 40~hours and reached the maximum value
of 20.4$\pm$3.9\%.
The target was then irradiated by a 71~MeV/nucleon $^6$He beam for 55~hours, 
by a 80~MeV/nucleon $^4$He beam for the following 25~hours, and again by
the $^6$He beam for 60~hours.
The magnitude of average polarization was found to be 13.8$\pm$2.7\%.
The target polarization slowly decreased as a function of time, which is
due to beam-irradiation damage in the target material.
This radiation damage increased the relaxation rate of the target
material from $\Gamma=$ 0.127(6)~h$^{-1}$ before the experiment
to $\Gamma=$
0.295(4)~h$^{-1}$ after the beam irradiation.
The direction of the target polarization was reversed three times
during the measurement to cancel spurious asymmetries.
The 180$^{\circ}$ pulse NMR method was used here.
Reversal efficiency of 60--70\% was achieved.

\section{Data Reduction} \label{sec_data}

\subsection{Data analysis}

In principle,
elastic-scattering events of the $^6$He from protons can be
identified by the coincidence detection of scattered $^6$He particles
and recoil
protons, since the $^6$He does not have a bound excited state.
Note that the first excited state of $^6$He, which is the $2^+$ state at
1.87~MeV, is above the two-neutron breakup threshold (0.975~MeV).
Thus, any excited $^6$He particles decay into
$\alpha$+$n$+$n$ systems before reaching the detectors.

\begin{figure}[htbp]
 \begin{center}
  \includegraphics[width=0.87\linewidth]{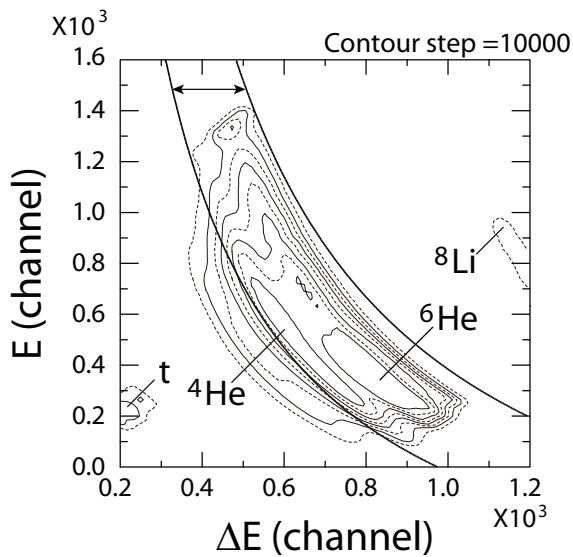}
  \caption{Two-dimensional plot of the total energies of scattered
  particles versus their energy losses. Solid curves indicate the
  particle-identification gate.}
  \label{fig:ede}
 \end{center}
\end{figure}

Scattered particles were identified by the standard $\Delta E$-$E$ method.
Figure~\ref{fig:ede} shows a two-dimensional plot of the total energies 
of scattered
particles $E$ versus their energy losses $\Delta E$, where 
loci of tritons, $^4$He, $^6$He, and $^8$Li are found.
Tritons and $^8$Li are the contamination in the secondary beam.
Most of $^4$He particles were produced by $^6$He dissociation in the
secondary target.
However, some originated from 
$^6$He reactions in the plastic scintillators.
So to count all of the $p+^6$He elastic-scattering events, the particle
identification gate includes most of the $^4$He locus as shown 
by solid curves in
Fig.~\ref{fig:ede}.
The contribution of the dissociation reaction, which is not excluded by this
gate, was subtracted using a kinematics relation. This is described 
after  the response of the recoil proton detectors is considered.

\begin{figure}[htbp]
 \begin{center}
  \includegraphics[width=0.84\linewidth]{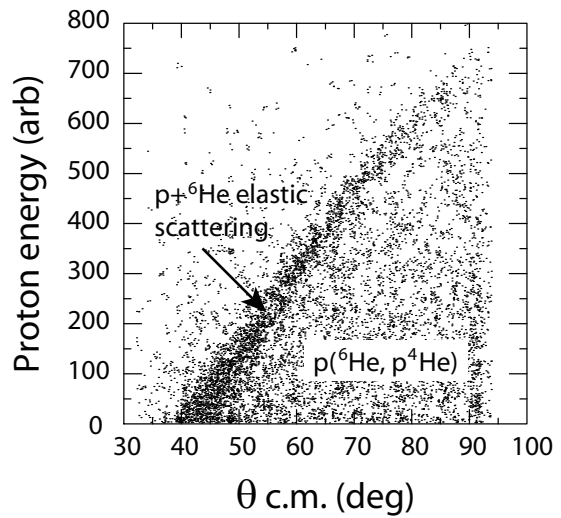}
  \caption{Two-dimensional plot of the proton energies versus their
  scattering angles.}
  \label{fig:csi}
 \end{center}
\end{figure}

Figure~\ref{fig:csi} shows a two-dimensional scatter plot of the proton energies versus
their scattering angles in the center-of-mass system, $\theta_{\textrm{c.m.}}$.
The kinematic locus of the elastic scattering is clearly identified, while
backgrounds 
from other reaction channels such as $p$($^6$He, $p^4$He) are also evident.
The kinematic locus of elastic-scattering events shows that the recoil
protons were properly detected outside of the target.
It should be noted that this correlation was not used for the event
selection, since it would cause a loss of the events at forward angles.

To discriminate elastic scattering from the background, 
we used the
correlation of the
azimuthal angles of protons $\phi_p$ with those of scattered particles
$\phi_{\textrm{scatt.}}$.
In the case of the elastic scattering, a scattered $^6$He and a recoil proton
stay within a well defined reaction plane 
since the final state is a binary system.
Thus,
the difference of
azimuthal angles 
$\Delta \phi=\phi_p-\phi_{\textrm{scatt.}}$ makes a narrow peak at
around 180$^{\circ}$.
This back-to-back correlation holds even if the scattered $^6$He is
dissociated in the plastic scintillator.
In the case of other reactions, however, the azimuthal angle difference 
is more spread
since their final states consist of more than two particles.
%

\begin{figure}[htbp]
 \begin{center}
  \includegraphics[width=0.75\linewidth]{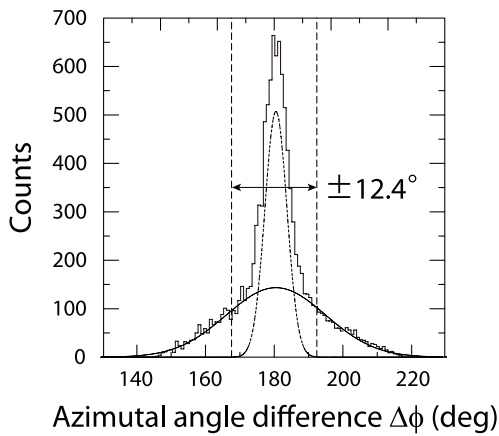}
  \caption{Azimuthal angle difference between scattered particles and
  recoil protons, fitted by a double-Gaussian function.}
  \label{fig:copl}
 \end{center}
\end{figure}

\begin{figure}[htbp]
 \begin{center}
  \includegraphics[width=0.99\linewidth]{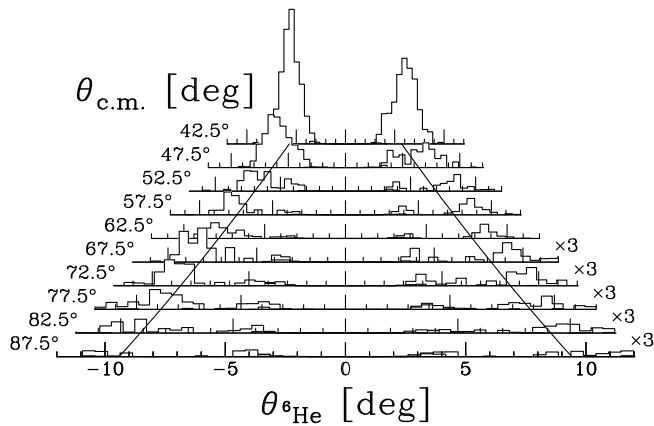}
  \caption{Scattering angle correlation between scattered particles and
  recoil protons. The solid curves indicate the kinematics of the
  $p+^6$He elastic scattering.}
  \label{fig:tcor}
 \end{center}
\end{figure}

Figure~\ref{fig:copl} shows the distribution of the azimuthal angle
difference $\Delta \phi$ 
fitted by a double-Gaussian function.
The narrower component is reasonably identified as that of the
elastic-scattering events.
The peak width of 3.5$^{\circ}$ in sigma is consistent with 
the detector resolution of 3.1$^{\circ}$.
We selected the events of $|\Delta \phi-180^{\circ}|<12.4^{\circ}$.
The background
remaining in the gate was 
evaluated from the broader component and was subtracted.
Contributions of the inelastic scattering and other reactions such as
breakup were removed in this way without losing the elastic-scattering
yields.
Figure~\ref{fig:tcor} shows a background-subtracted two-dimensional plot
of scattering
angles in the center-of-mass system versus angles of scattered particles.
Center-of-mass scattering angles were deduced from recoil angles of
the protons in the laboratory system,
since the resolution of scattering angles of $^6$He particles is
insufficient due to the kinematic focusing.
In Fig.~\ref{fig:tcor},
clear peaks of elastic-scattering events lie along the solid curves
indicating the kinematics of the $p+^6$He elastic scattering.
Small peaks at $|\theta_{^6\textrm{He}}|\approx 4^{\circ}$ originated 
from the ambiguity in the background subtraction.
Yields of the $p+^6$He elastic scattering were obtained by counting
the events of the elastic-scattering peaks 
in the typical width of $4^{\circ}$ in $\theta_{^6\textrm{He}}$.

The present work demonstrates the applicability of the solid 
polarized proton target in the RI-beam experiment.
The relaxed operation condition of the target, i.e. a low magnetic 
field of 0.1~T and high magnetic field of 100~K, enables us to
detect the low-energy recoil protons.
As described in the data analysis above, information on the trajectory 
of recoil proton is indispensable both in identifying the 
elastic-scattering events (Fig.~\ref{fig:copl}) and in deducing the
scattering angle (Fig.~\ref{fig:tcor}).

\subsection{Experimental data}

The $d\sigma/d\Omega$ of the $p+^6$He elastic
scattering measured at 71~MeV/nucleon are summarized in 
Table~\ref{table:dcsdata}.
In the backward region, the uncertainty mainly results from statistics
and from the ambiguity in the background subtraction.
In the forward angular region, $\theta_{\textrm{c.m.}}<60^{\circ}$, the main
component of the uncertainty in $d\sigma/d\Omega$ is the
systematic uncertainty in the 
number of the incident particles (10\%).
The target was hit by only a fraction of beam particles since 
the size of the secondary beam was comparable to that of the target.
The percentage of the beam particles incident on the target was
determined from the beam profile and was found to
be 65$\pm$7\% of those counted by the beam monitor. 
The beam profile was measured with the MWDC by removing the
beam stopper.
Stability of the beam profile was confirmed by several
measurements carried out before, during, and after the
elastic-scattering measurement.

The analyzing power $A_y$ is deduced with the standard
procedure as 
\begin{equation}
 A_y =\frac{1}{P}\frac{L-R}{L+R}, \ \ \nonumber
  L=\sqrt{N_L^{\uparrow}\cdot N_R^{\downarrow}}, \ \ \nonumber
  R=\sqrt{N_L^{\downarrow}\cdot N_R^{\uparrow}}, \nonumber
\end{equation}
where $P$ denotes the target polarization.
The values $N$'s represent the yield of the elastic-scattering events where
subscripts and superscripts denote the scattering direction (left/right) and
the polarization direction (up/down), respectively.
The statistical uncertainty is expressed by 
\begin{eqnarray}
& & \frac{\Delta A_y }{A_y} =\frac{LR}{L^2-R^2}\ 
\sqrt{\frac{1}{N_R^{\uparrow}}+\frac{1}{N_R^{\downarrow}}+\frac{1}{N_L^{\uparrow}}+\frac{1}{N_L^{\downarrow}}}. \nonumber
\end{eqnarray}
This procedure allows us to minimize the systematic uncertainties
originating from unbalanced detection efficiencies and misalignment of
detectors.
The obtained $A_y$ are summarized in Table~\ref{table:aydata}.
It must be noted that there is an additional scale error of 19\%
resulting from the uncertainty in the target polarization $P$ (see
Appendix.~\ref{sec:calib}).

\begin{table}[htbp]
\begin{center}
\begin{tabular}{cc|cc}
\hline
\hline
$\theta_{\textrm{c.m.}}$ (deg) & $\Delta \theta_{\textrm{c.m.}}$ (deg) & $\frac{d\sigma}{d\Omega}$ (mb/sr) & $\Delta \frac{d\sigma}{d\Omega}$ (mb/sr) \\
\hline
42.1 & 2.5  & 5.02\phantom{0}  & 0.52\phantom{0}  \\
47.1 & 2.5  & 2.03\phantom{0}  & 0.22\phantom{0}  \\
52.1 & 2.5  & 0.796  & 0.098  \\
57.4 & 2.5  & 0.454  & 0.059  \\
62.3 & 2.5  & 0.360  & 0.046  \\
67.3 & 2.5  & 0.226  & 0.031  \\
72.3 & 2.5  & 0.172  & 0.023  \\
77.3 & 2.5  & 0.127  & 0.018  \\
82.2 & 2.5  & 0.064  & 0.013  \\
87.2 & 2.5  & 0.038  & 0.012  \\
\hline
\hline
\end{tabular}
\end{center}
\caption{Differential cross sections for the $p+^6$He elastic scattering
 at 71~MeV/nucleon. The $\Delta \theta_{\textrm{c.m.}}$ denotes the bin width. 
The
 $\Delta \frac{d\sigma}{d\Omega}$ denotes the quadratic sum 
 of the statistical and systematic uncertainties. 
}
\label{table:dcsdata}
\end{table}

\begin{table}[htbp]
\begin{center}
\begin{tabular}{cc|cc}
\hline
\hline
$\theta_{\textrm{c.m.}}$ (deg) & $\Delta \theta_{\textrm{c.m.}}$ (deg) & \ \ \ \ \ \ $A_y$ \ \ \ \ \ \ & \ \ \ \ \ \
 $\Delta A_y$ \ \ \ \ \ \ \\
\hline
37.1 &  2.5  &   \phantom{$-$}0.242  &  0.069 \\
44.6 &  5.0  &   \phantom{$-$}0.021  &  0.089 \\
54.6 &  5.0  &  $-$0.016  &  0.135 \\
64.8 &  5.0  &   \phantom{$-$}0.11\phantom{$0$}  &  0.18\phantom{$0$} \\
74.3 &  5.0  &  $-$0.27\phantom{$0$}  &  0.27\phantom{$0$} \\
\hline
\hline
\end{tabular}
\end{center}
\caption{Analyzing powers for the $p+^6$He elastic scattering at
 71~MeV/nucleon. 
$\Delta A_y$ denotes the statistical uncertainty.
Note that there is an additional scale error of
 19\% resulting from the uncertainty in the target polarization. 
 The $\Delta \theta_{\textrm{c.m.}}$ denotes the bin
 width.}
\label{table:aydata}
\end{table}

\begin{figure}[htbp]
 \begin{center}
  \includegraphics[width=0.87\linewidth]{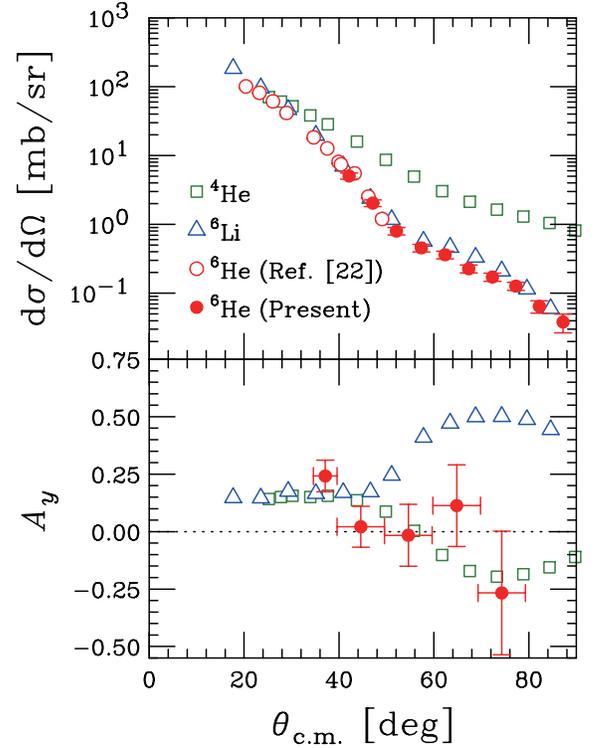}
  \caption{(Color online) Differential cross sections and analyzing
  powers of the 
  $p+^4$He at 72~MeV (open squares: Ref.~\cite{Burzynski89}), of the 
  $p+^6$Li at 72~MeV (open triangles: Ref.~\cite{Henneck94}), and of the 
  $p+^6$He at 71~MeV (open circles: Ref.~\cite{Korsheninnikov97}, closed
  circles: present work).}
  \label{fig:dcsay}
 \end{center}
\end{figure}

Figure~\ref{fig:dcsay} shows $d\sigma/d\Omega$ and $A_y$ for 
the $p+^6$He elastic scattering at 71~MeV/nucleon (closed circles: present work,
open circles: Ref.~\cite{Korsheninnikov97}), 
those for the $p+^4$He at 72~MeV/nucleon (open squares:
Ref.~\cite{Burzynski89}), and
those for the $p+^6$Li at 72~MeV/nucleon (open triangles:
Ref.~\cite{Henneck94}).
The present data are
consistent with the previous ones in Ref.~\cite{Korsheninnikov97} in an
overlapping
angular region of $\theta_{\textrm{c.m.}}=40$--$50^{\circ}$.
We extended the data to the backward angles of 
$\theta_{\textrm{c.m.}}\approx 90^{\circ}$.
It is found that the $d\sigma/d\Omega$ of $p+^6$He are almost identical
with those of $p+^6$Li at $\theta_{\textrm{c.m.}}=20$--$90^{\circ}$, while
they have a 
steeper angular dependence than those of $p+^4$He.
In good contrast to the similarity found in $d\sigma/d\Omega$, $A_y$
data are widely different between $p+^6$He and $p+^6$Li.
The $A_y$ of $p+^6$Li increase as a function of the scattering angle in
an angular region of $\theta_{\textrm{c.m.}}=40$--$70^{\circ}$ and take 
large positive values.
This behavior is commonly seen in proton elastic scattering from stable
nuclei at 
the present energy region~\cite{Sakaguchi82}.
Unlike this global trend, $A_y$ of $p+^6$He decreases in 
$\theta_{\textrm{c.m.}}=37$--$55^{\circ}$, which is rather similar to those
of $p+^4$He.
While the large error bars prevent us from observing the difference
between $A_y$ of $p+^6$He and of $p+^4$He, it is clearly seen that the
angular distribution of the $A_y$ in 
$p+^6$He deviates from that of $p+^6$Li.

\section{Phenomenological Optical Model Analysis} \label{sc:phen}

\subsection{Optical potential fitting}

The aim of this section is to extract the gross characteristics of the
spin-orbit interaction between a proton and $^6$He.
For this purpose, we determined the optical model potential that
reproduces the experimental data of both differential cross sections and
analyzing powers.
The optical model potential obtained in this phenomenological approach
will be compared with the semi-microscopic calculations in
Section~\ref{sec:micro}.

We adopted a standard Woods-Saxon optical potential with a spin-orbit
term of the Thomas form:
\begin{eqnarray}
U_{\textrm{OM}}(R) =  &-&V_0 \, f_r (R) - i \, W_0 \, f_i (R)  \nonumber
 \\
                      &+& 4i\, a_{id} \, W_d \, \frac{d}{dR} f_{id}(R)  \nonumber \\
      &+&  V_s \, \frac{2}{R} \, \frac{d}{dR} f_s (R) \; \bm{L} \cdot \bm{\sigma}_p
          + V_{\textrm{C}} (R)
\label{eq:op1}
\end{eqnarray}
with
\begin{eqnarray}
f_x (R) &=& \left[ 1 + \exp \left( \frac{R-r_{0x} A^{1/3}}{a_x}
                        \right) \right] ^{-1} \; \\ 
(x&=&r,i,id,\ \textrm{or}\ s).\nonumber 
\label{eq:op2}
\end{eqnarray}
Here, $\bm{R}$ is the relative coordinate between a proton and a
${}^{6}$He particle (see Fig.~\ref{fig:coord} (b)),
$\bm{L} = \bm{R}\times(-i\hbar\bm{\nabla}_{R})$
is the associated angular momentum, and $\bm{\sigma}_{p}$ is the Pauli
spin operator of the proton.
The subscripts $r$, $i$, $id$, and $s$ denote real, volume imaginary,
surface imaginary, and spin-orbit, respectively.
$V_{\textrm{C}}$ is the Coulomb potential of uniformly charged
sphere with a radius of 
$r_{0\textrm{C}} A ^{1/3}$ fm ($r_{0\textrm{C}}=1.3$ fm).

The search procedure for the best-fit potential parameters was 
made in two 
steps: first the parameters of the central term were found by minimizing the
$\chi^2$ values of $d\sigma/d\Omega$, and second the parameters of the
spin-orbit term by fitting $A_y$.
These two steps were iterated alternately until 
convergence was achieved.
Such a procedure is feasible since the contribution of the spin-orbit
potential to $d\sigma/d\Omega$ is much smaller than those of the central
terms. 
In the fitting, 
we used the data in 
Ref.~\cite{Korsheninnikov97} and the present ones.
Uncertainties of $d\sigma/d\Omega$ smaller than 10\% were
artificially set to 10\% in order to avoid trapping in an unphysical
local $\chi^2$ minimum. 
The fitting was carried out using the ECIS79 code~\cite{Raynal65}.
A set of parameters for the $p+^6$Li elastic scattering at
72~MeV/nucleon~\cite{Henneck94}, labeled as Set-A in Table~\ref{table:param},
was used as the initial values in the search of the $p$-$^6$He potential
parameters.

\begin{table*}[htbp]
\caption{Parameters of the optical potentials for $p+^6$Li
at 72~MeV/nucleon~\cite{Henneck94} and $p+^6$He at 71~MeV/nucleon (\cite{Gupta00}
 and present work).}
\begin{ruledtabular}
\begin{tabular}{ll|cccccccccccc}
 &   &  $V_0$  &  $r_{0r}$  &  $a_{r}$  &  $W_0$  &  $r_{0i}$ &  $a_{i}$
   &  $W_d$  &  $r_{0id}$ &  $a_{id}$
   & $V_{s}$ & $r_{0s}$ & $a_{s}$ \\
 &   &  (MeV)  &  (fm)  &  (fm)  &  (MeV)  &  (fm)  &  (fm)
   &  (MeV)  &  (fm)  &  (fm)
   &  (MeV)  &  (fm)  &  (fm) \\ \hline
Set-A & $p+^6$Li \cite{Henneck94} & 31.67 & 1.10 & 0.75 & 14.14 & 1.15 & 0.56
      & --- & ---  & --- & 3.36 & 0.90 & 0.94 \\
Set-B & $p+^6$He (Present) & 27.86 & 1.074 & 0.681 & 16.58 & 0.86 & 0.735
      & --- & ---  & --- & 2.02 & 1.29 & 0.76 \\
Set-C & $p+^6$He \cite{Gupta00} & 30.00 & 0.990 & 0.612 & 14.0 & 1.10 & 0.690
      & 1.00 & 1.76 & 0.772 & 5.90 & 0.677 & 0.630 \\
\end{tabular}
\end{ruledtabular}
\label{table:param}
\end{table*}

\begin{figure}[htbp]
 \begin{center}
  \includegraphics[width=0.8\linewidth]{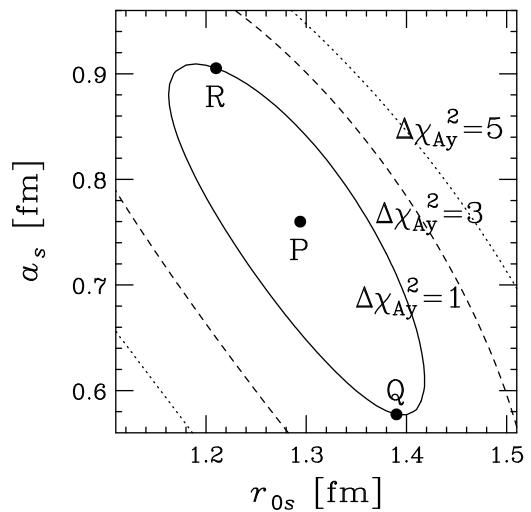}
  \caption{
Contour map of the $\Delta\chi^2_{A_y}$ values (see the text for the
 definition) on the two-dimensional plane of $r_{0s}$ and $a_s$.
The solid, dashed, and dotted curves indicate $\Delta\chi^2_{A_y}$=1,
 3, and 5, respectively.
The point that gives the best-fit parameters, Set-B in Table IV, is
 indicated by the point-P.
See the text for the points-Q and -R.
}
  \label{fig:cont}
 \end{center}
\end{figure}

The parameters obtained for the $p+{}^6$He elastic scattering are
labeled as Set-B in Table~\ref{table:param}.
The reduced $\chi^2$ values for $d\sigma/d\Omega$ and $A_y$ were 0.95 and
0.96, respectively.
Uncertainties of the parameters of the spin-orbit potential, $r_{0s}$,
$a_s$, and $V_s$, are evaluated in the following manner.
Figure~\ref{fig:cont} shows the contour map of the deviation of $\chi^2$ value for
$A_y$ from that calculated by the Set-B (as indicated by the point-P),
$\Delta\chi^2_{A_y}$, on the two-dimensional plane of $r_{0s}$ and $a_s$
after projecting with optimized $V_s$ at each point of the plane.
In the figure, a simultaneous confidence region for $r_{0s}$ and $a_s$
is presented by the solid contour indicating $\Delta\chi^2_{A_y} = 1$.
In this region, the optimum $V_s$ ranges between 1.15~MeV (at the
point-Q) and 2.82~MeV (at the point-R).
In the $r_{0s}$-$a_s$-$V_s$ space, a surface that $\Delta\chi^2_{A_y} = 1$
touches planes that are expressed by $r_{0s} = 1.29 \pm 0.13$ fm,
$a_s = 0.76 \pm 0.17$ fm, and $V_s = 2.02 \pm 0.87$~MeV, which
gives a rough estimation for uncertainties of the parameters.

\subsection{Characteristics of spin-orbit potential}

\begin{figure}[htbp]
 \begin{center}
  \includegraphics[width=0.87\linewidth]{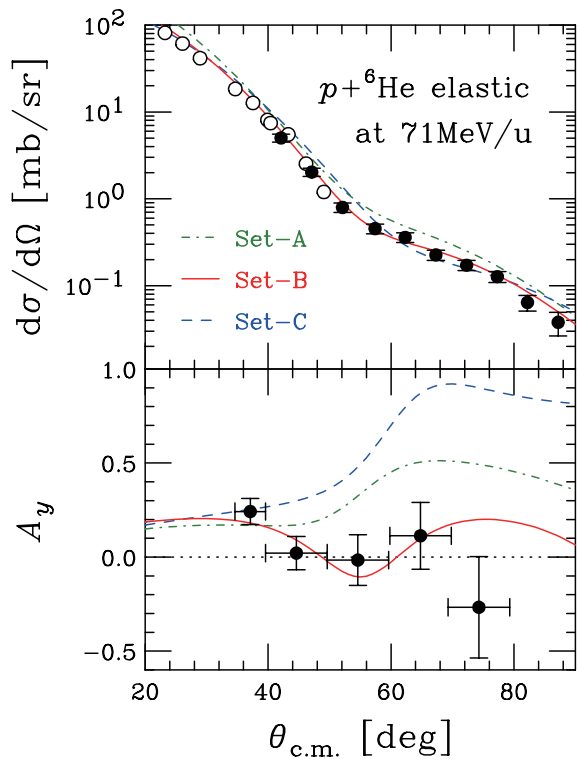}
  \caption{(Color online) Differential cross sections
  and analyzing powers calculated by the phenomenological optical potentials
  are shown together with the experimental data.
  The dot-dashed curves denote calculations of the Set-A in
  Table~\ref{table:param},
  the solid curves those of set-B, and the dashed curves those of Set-C.
  Solid circles are present data and open circles are from
  Ref.~\cite{Korsheninnikov97}.
  }
  \label{fig:phen}
 \end{center}
\end{figure}

\begin{figure}[htbp]
 \begin{center}
  \includegraphics[width=0.82\linewidth]{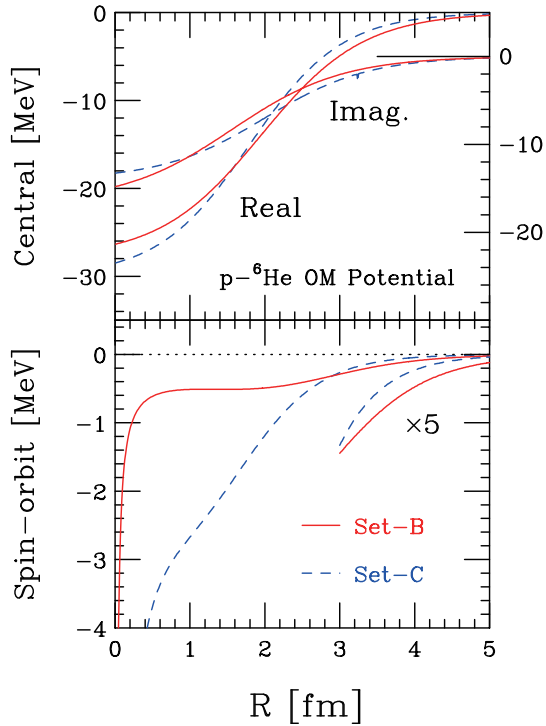}
  \caption{(Color online) Radial dependences of the phenomenological
  optical potential (Set-B and Set-C in Table~\ref{table:param}). Solid and
  dashed curves in the upper panel represent the real and imaginary
  parts of the central term. The lower panel shows the spin-orbit
  potential.}
  \label{fig:potcomp}
 \end{center}
\end{figure}

In Fig.~\ref{fig:phen}, the results of calculations of the observables 
made with the optical
potentials of Set-A,
-B, and -C in Table~\ref{table:param} are shown together with the
experimental data.
Set-C was taken from Ref.~\cite{Gupta00}, where a phenomenological
optical model potential that reproduced only the previous
$d\sigma/d\Omega$ data of the $p+^6$He at
71~MeV/nucleon~\cite{Korsheninnikov97} was reported.
The radial dependences of the  
$p$-$^6$He optical potentials (Set-B and Set-C) 
are shown in 
Fig.~\ref{fig:potcomp} by solid and dashed lines, respectively.

The calculation with the potential Set-C reasonably reproduces the
present $d\sigma/d\Omega$ data, whereas it largely deviates from the
$A_y$ data at $\theta_{\textrm{c.m.}} \agt 40^{\circ}$.
It should be noted that the $A_y$ data were unavailable when the
potential Set-C was sought.
The calculation with the potential Set-B reproduces both
$d\sigma/d\Omega$ 
and $A_y$ over whole angular region 
except for the most backward data point of $A_y$.
Similarity of the $d\sigma/d\Omega$ calculated with Set-B and Set-C
potentials originates from that of the central terms as shown in the
upper panel of Fig.~\ref{fig:potcomp}.
The reliability of the potential obtained in the present work is supported
by the fact that two independent analyses yielded the similar results
for the central terms.
In contrast to the central terms, the spin-orbit terms of these two
potentials are quite different, resulting in a large difference in $A_y$
as shown in Fig.~\ref{fig:phen}.
Note that the present data are sensitive to the optical potential in a
region of $R\agt$1.5~fm.
The spin-orbit potential of Set-B is much shallower than that of Set-C
at $R\alt$ 2.8~fm, while it is deeper at larger radii.
This is due to the small value of $V_s$ and large values of $r_{0s}$ and
$a_s$ of Set-B compared with those of Set-C.
The phenomenological optical model analysis suggests that 
the $A_y$ data can be reproduced only
with a shallow and long-ranged spin-orbit potential.

The parameters of our spin-orbit potential are compared 
with those of neighboring even-even stable
nuclei and with global potentials in Table~\ref{table:lsparam}.
Phenomenological optical potentials for the 
$p+^{16}$O at 65~MeV and $p+^{12}$C at 16--40~MeV
are taken from Ref.~\cite{Sakaguchi82} and Refs.~\cite{Fricke67,Fabrici80},
respectively.
In addition to these local potentials, we also examined the parameters
of global optical potentials: CH89~\cite{Varner91} and
Koning-Delaroche (KD)~\cite{Koning03}, of which applicable ranges are $E=$
10--65~MeV, $A=$ 40--209
and $E=$ 0.001--200~MeV, $A=$ 24--209, respectively.
While they are constructed for nuclei heavier than $^6$He, it is worthwhile
comparing them, since the mass-number dependence of the parameters
is relatively small.
For example, the mass-number dependence appears only in $r_{0s}$ in the
case of CH89~\cite{Varner91} as: 
\begin{eqnarray}
 V_s &=& 5.9(1)~\textrm{MeV}, \nonumber \\
 r_{0s} &=& 1.34(3) - 1.2(1) A^{-1/3}~\textrm{fm}, \nonumber \\
 a_s &=& 0.63(2)~\textrm{fm}.\nonumber
\end{eqnarray}
Table~\ref{table:lsparam} includes the parameters of these potentials
for the nuclei within the applicable range.
Incident energies of $E=$~65~MeV and $E=$~71~MeV were assumed here for CH89
and KD, respectively.

\begin{table*}[htbp]
\begin{center}
\begin{tabular}{l|lll}
\hline
\hline
   & \hspace{5mm} $V_{s}$ (MeV) & $r_{0s}$ (fm) & $a_{s}$ (fm) \\
\hline
\hline
$p+^6$He, $E=$ 71~MeV (Set-B) & \hspace{5mm} 2.02 & 1.29 & 0.76 \\
$p+^{12}$C, $E=$ 40~MeV~\cite{Fricke67} & \hspace{5mm} 6.18 & 1.109 &  0.517  \\
$p+^{12}$C, $E=$ 16--40~MeV~\cite{Fabrici80} & \hspace{5mm} 6.4 & 1.00 &  0.575  \\
$p+^{16}$O, $E=$ 65~MeV~\cite{Sakaguchi82} & \hspace{5mm} 5.793 & 1.057 & 0.5807 \\
\hline
CH89, $E=$ 65~MeV, $A=$40--209~\cite{Varner91} \hspace{5mm} & \hspace{5mm} 5.9$\pm$0.1 & 0.99--1.14 & 0.63$\pm$0.02 \\
KD, $E=$ 71~MeV, $A=$24--209~\cite{Koning03} &  \hspace{5mm} 4.369--4.822  \hspace{5mm} &  0.961--1.076  \hspace{5mm} & 0.59 \\
\hline
\hline
\end{tabular}
\end{center}
\caption{Parameters of the spin-orbit term of phenomenological and
 global optical potentials.}
\label{table:lsparam}
\end{table*}

\begin{figure}[htbp]
 \begin{center}
  \includegraphics[width=0.95\linewidth]{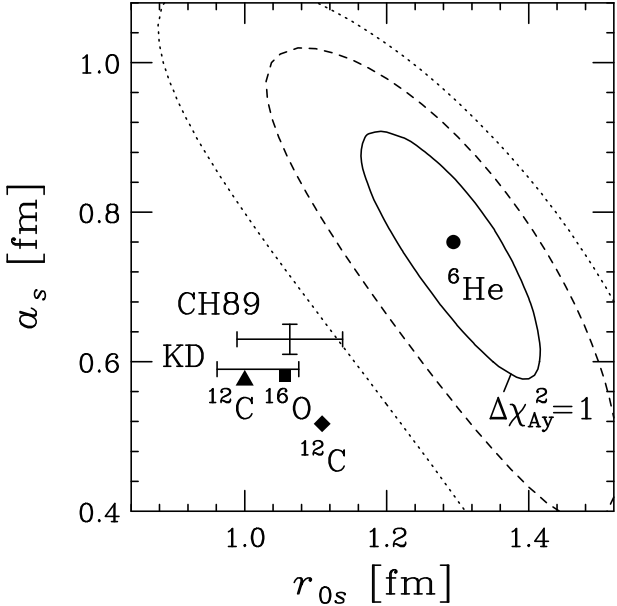}
  \caption{
  Two-dimensional distribution of $r_{0s}$ and $a_s$ of
  phenomenological OM potentials for $p+^6$He
  (closed circle), $p+^{12}$C (closed triangle: Ref.~\cite{Fabrici80}, closed diamond:~\cite{Fricke67}), and
  $p+^{16}$O (closed square:~\cite{Sakaguchi82}).
  The solid contour indicates the simultaneous confidence region for the
  $r_{0s}$ and $a_s$ values for the $p+^6$He as displayed in
  Fig.~\ref{fig:cont}.
  Parameters of global OM
  potentials~\cite{Varner91,Koning03} are also shown by solid lines which represent the
  $A$-dependence of $r_{0s}$. 
  }
  \label{fig:ct}
 \end{center}
\end{figure}

Firstly, we focus on $r_{0s}$ and $a_s$ to discuss the radial shape of
the spin-orbit potential.
Combination of different values of $r_{0s}$ and $a_s$ can provide similar
results of
$A_y$ since the
observable is sensitive to the surface region of the spin-orbit
potential.
We thus compare these parameters on 
the two-dimensional plane of $r_{0s}$ and $a_s$ as shown in
Fig.~\ref{fig:ct}.
Parameters for the stable nuclei are mostly distributed in a region of
$r_{0s}=0.8$--$1.1$~fm and $a_s=0.5$--$0.6$~fm, whereas 
that for $^6$He is located in the upper right side of the
figure.
These large $r_{\textrm{0}s}$ and/or $a_s$ values indicate that the
spin-orbit potential between a proton and a $^6$He has a long-ranged
nature compared with those for stable nuclei.
The depth parameter $V_s$ was also compared with the global
systematics.
The $V_s$ value of $p$-$^6$He potential was found to be 2.02~MeV for the
best-fit potential (Set-B) and ranges between 1.15 and 2.82~MeV in the
simultaneous confidence region for $r_{0s}$ and $a_s$.
On the other hand, those of stable nuclei are mostly distributed around 5~MeV
as shown in Table~\ref{table:lsparam}.
Comparing these values, the depth parameter of the spin-orbit potential
between a proton and a
$^6$He is found to be much smaller than those of stable nuclei.

The phenomenological analysis indicates that
the spin-orbit potential between a proton and $^6$He is characterized 
by large $r_{0s}/a_s$ and small $V_s$ values yielding shallow and
long-ranged radial dependence.
Intuitively, these characteristics 
can be understood 
from the diffused density distribution of $^6$He.
However, its microscopic origin can not be clarified by the
phenomenological approach.
To examine the microscopic origin of the characteristics of the
$p$-$^6$He interaction,
microscopic and semi-microscopic analyses are required.
Section~\ref{sec:micro} describes one of such analyses based on a
cluster folding model for $^6$He.

\section{Semi-microscopic Analyses} \label{sec:micro}


In this section, 
we examine two kinds of the folding potential, 
the cluster folding (CF) and the nucleon folding (NF) ones. 
They are compared with
the phenomenological optical model (OM) potential determined 
in the preceding section.
The results of calculations of observables made by these potentials are
compared with the experimental data.

In the CF potential, we adopt the $\alpha n n$ cluster model for $^6$He and
fold interactions between the proton and the valence neutrons, $V_{pn}$,
with the neutron density in $^6$He and those between the proton and the
$\alpha$ core,
$V_{p \alpha}$, with the $\alpha$ density in $^6$He.
In the NF potential, we decompose the $\alpha$ core into two neutrons and two protons
and fold the interactions between the incident proton and the four neutrons, $V_{pn}$,
with the neutron density in $^6$He and those between the incident proton and
two target protons, $V_{pp}$, with the proton density in $^6$He.

The detailed expressions of such folding potentials are given in the
following subsection,
where the Coulomb interaction is considered in the $p$-$p$ and
$p$-$\alpha$ interactions respectively 
when compared with the corresponding scattering data but finally it is assumed
to act between the proton and the $^6$He target with $r_{0\textrm{C}}=
1.400$ fm~\cite{Gupta00}.

\subsection{\label{sec:level2}Folding potentials}

Denoting two valence neutrons by $n_1$ and $n_2$,
the CF potential $U_{\text{CF}}$ is given as
\begin{eqnarray}
U_{\textrm{CF}} &=& \int V_{p n_1} \, \rho_n^{\textrm{CF}} (r_1) \, d\bm{r}_1
      + \int V_{p n_2} \, \rho_n^{\textrm{CF}} (r_2) \, d\bm{r}_2\nonumber \\
      &+& \int V_{p \alpha} \, \rho_\alpha^{\textrm{CF}} (r_\alpha) \, d\bm{r}_\alpha \;,
\label{eq:ucf1}
\end{eqnarray}
where $\bm{r}_1$, $\bm{r}_2$, and $\bm{r}_\alpha$ are the position vectors of $n_1$, $n_2$, and
the $\alpha$ core from the center of mass of $^6$He, respectively.
The neutron and $\alpha$ densities, $\rho_n^{\textrm{CF}}$ and $\rho_\alpha^{\textrm{CF}}$, are calculated by
the $\alpha n n$ cluster model for $^6$He~\cite{Hiyama96,Hiyama03},
where the condition $\bm{r}_1 + \bm{r}_2 + 4 \bm{r}_\alpha = 0$ is considered as usual.

\begin{figure}[tb]
\includegraphics[width=0.70\linewidth,clip]{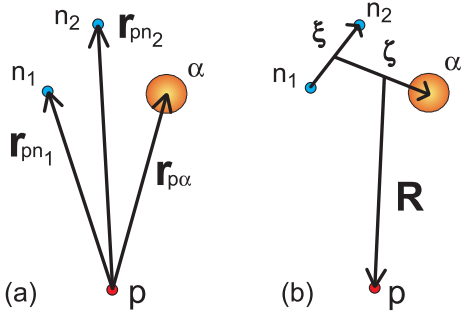}
\caption{(Color online) Coordinate systems for the cluster folding model.}
\label{fig:coord}
 \end{figure}

In the present work, we specify the potentials in the right hand side of
Eq.~(\ref{eq:ucf1}) by the central plus spin-orbit (LS) type:
\begin{equation}
V_{pn_i} = V^0_{pn}(\vert\bm{r}_{pn_i}\vert) +
V^\text{LS}_{pn}(\vert\bm{r}_{pn_i}\vert)
\bm{\ell}_{pn_i}\cdot(\bm{\sigma}_p+\bm{\sigma}_{n_i}), \nonumber
\end{equation}
where $i=1,2$ and
\begin{equation}
V_{p\alpha}=V^0_{p\alpha}(\vert\bm{r}_{p\alpha}\vert ) +
V^\text{LS}_{p\alpha}(\vert\bm{r}_{p\alpha} \vert ) 
 \bm{\ell}_{p\alpha} \cdot \bm{\sigma}_{p}.
\label{eq:vp}
\end{equation}
Here, $\bm{r}_{pn_1}$, $\bm{r}_{pn_2}$, and $\bm{r}_{p\alpha}$ are
defined in Fig.~\ref{fig:coord} (a), and 
$\bm{\ell}_{pn_1} =
\bm{r}_{pn_1}\times(-i\hbar\bm{\nabla}_{pn_1})$, etc.

In the following, we transform the set of coordinates 
$(\bm{r}_{pn_1}, \bm{r}_{pn_2}, \bm{r}_{p\alpha})$ 
to that of 
$(\bm{\xi}, \bm{\zeta}, \bm{R})$, which are defined in
Fig.~\ref{fig:coord} (b), to
describe the angular momenta $\bm{\ell}_{pn_i}$ and
$\bm{\ell}_{p\alpha}$ in terms of $\bm{L}$.
The transformation is  
\begin{eqnarray}
\bm{r}_{p n_1} &=& - \bm{R} - \frac{2}{3} \bm{\zeta} - \frac{1}{2} \bm{\xi} \;, \;\;\;
\bm{r}_{p n_2} = - \bm{R} - \frac{2}{3} \bm{\zeta} + \frac{1}{2} \bm{\xi} \;, \;\;\;\nonumber\\
\bm{r}_{p \alpha} &=& - \bm{R} + \frac{1}{3} \bm{\zeta} \;,
\label{eq:coor}
\end{eqnarray}
and consequently 
\begin{eqnarray}
\bm{\nabla}_{p n_1} &=& - \frac{1}{6} \bm{\nabla}_R - \frac{1}{2} \bm{\nabla}_\zeta - \bm{\nabla}_\xi \;, \;\;\;\nonumber\\
\bm{\nabla}_{p n_2} &=& - \frac{1}{6} \bm{\nabla}_R - \frac{1}{2} \bm{\nabla}_\zeta + \bm{\nabla}_\xi \;, \;\;\;\nonumber\\
\bm{\nabla}_{p \alpha} &=& - \frac{2}{3} \bm{\nabla}_R + \bm{\nabla}_\zeta \;.
\label{eq:nab}
\end{eqnarray}

These relations lead to, for example, 
\begin{equation}
\bm{\ell}_{p n_1} = (- \bm{R} - \frac{2}{3} \bm{\zeta} - \frac{1}{2} \bm{\xi}) \times
   [ -i \hbar (- \frac{1}{6} \bm{\nabla}_R - \frac{1}{2} \bm{\nabla}_\zeta - \bm{\nabla}_\xi) ] \;.
\label{eq:ell1}
\end{equation}
Here, $\bm{\nabla}_{\xi}$ and $\bm{\nabla}_{\zeta}$ can be neglected, because these are the momenta for the internal degrees of freedom of ${}^6$He and their expectation values are zero for a spherically symmetric nucleus~\cite{Rikus84}. 
Using $\frac23 \bm{\xi} + \frac12\bm{\zeta} = - \bm{r}_1$, we get
\begin{equation}
\bm{\ell}_{p n_1} = \frac{1}{6} \, [ \bm{L} - \bm{r}_1 \times (-i \hbar \bm{\nabla}_R) ] \;,
\label{eq:ell2}
\end{equation}
which is independent of the special choice of the $^6$He internal coordinates,
$\bm{\xi}$ and $\bm{\zeta}$.
To $\bm{L}$, $\bm{r}_1$ can contribute by its component along the $\bm{R}$ direction~\cite{Rikus84},
then
\begin{equation}
\bm{\ell}_{p n_1} = \frac{1}{6} \, \bm{L} \, ( 1 - \frac{\bm{r}_1 \cdot \bm{R}}{R^2} ) \;.
\label{eq:ell3}
\end{equation}
Similar expressions are obtained for $\bm{\ell}_{p n_2}$ and $\bm{\ell}_{p \alpha}$.
Setting $\frac{1}{2} (\bm{\sigma}_{n_1} + \bm{\sigma}_{n_2} ) = 0$ and
considering other quantities to appear in symmetric manners on 1 and 2,
we obtain the $p$-$^6$He potential as
\begin{equation}
U_{\textrm{CF}} = U_0^{\textrm{CF}}(R) + U_{\textrm{LS}}^{\textrm{CF}}(R) \, \bm{L} \cdot \bm{\sigma}_p  \;,
\label{eq:ucf2}
\end{equation}
with
\begin{eqnarray}
U_0^{\textrm{CF}}(R) &=& 2 \, \int V_{pn}^0 (|\bm{r}_1 - \bm{R}| ) \, \rho _n^{\textrm{CF}} (r_1) \, d \bm{r}_1\nonumber\\
  &+& \int V_{p \alpha}^0 (|\bm{r}_\alpha - \bm{R}| ) \, \rho_\alpha^{\textrm{CF}} (r_\alpha) \, d \bm{r}_\alpha
\label{eq:ucf0}
\end{eqnarray}
and
\begin{eqnarray}
U_{\textrm{LS}}^{\textrm{CF}}(R) &=& \frac{1}{3} \, \int V_{pn}^{\textrm{LS}} (|\bm{r}_1 - \bm{R}| ) \, 
           \left\{ 1 - \frac{\bm{r}_1 \cdot \bm{R}}{R^2} \right\} \, \rho _n^{\textrm{CF}} (r_1) \, d \bm{r}_1
               \nonumber \\
  &+& \frac{2}{3} \, \int V_{p \alpha}^{\textrm{LS}} (|\bm{r}_\alpha - \bm{R}| ) \, 
           \left\{ 1 - \frac{\bm{r}_\alpha \cdot \bm{R}}{R^2} \right\} \,
           \rho_\alpha^{\textrm{CF}} (r_\alpha) \, d \bm{r}_\alpha \;.
\label{eq:ucfls}
\end{eqnarray}

In a way similar to the above development, we get the NF model potential
$U_{\textrm{NF}}$.
In this case, the relative coordinates between the incident proton and
six nucleons in the $^6$He nucleus are transformed to the proton-$^6$He
relative coordinate $\bm{R}$ and a set of five independent internal
coordinates of $^6$He.
The obtained $U_{\textrm{NF}}$, which is independent on the choice of
the set of the internal coordinates, is written as
\begin{equation}
U_{\textrm{NF}} = U_0^{\textrm{NF}}(R) + U_{\textrm{LS}}^{\textrm{NF}}(R) \, \bm{L} \cdot \bm{\sigma}_p  \;,
\label{eq:u6bf2}
\end{equation}
with
\begin{eqnarray}
U_0^{\textrm{NF}}(R) &=& 2 \, \int V_{pp}^0 (|\bm{r}_1 - \bm{R}| ) \, \rho _p^{\textrm{NF}} (r_1) \, d \bm{r}_1\nonumber\\
       &+& 4 \, \int V_{pn}^0 (|\bm{r}_2 - \bm{R}| ) \, \rho _n^{\textrm{NF}} (r_2) \, d \bm{r}_2
\label{eq:u6bf0}
\end{eqnarray}
and
\begin{eqnarray}
U_{\textrm{LS}}^{\textrm{NF}}(R) &=&  \frac{1}{3} \, \int V_{pp}^{\textrm{LS}} (|\bm{r}_1 - \bm{R}| ) \, 
           \left\{ 1 - \frac{\bm{r}_1 \cdot \bm{R}}{R^2} \right\} \, \rho _p^{\textrm{NF}} (r_1) \, d \bm{r}_1
               \nonumber \\
      &+& \frac{2}{3} \, \int V_{pn}^{\textrm{LS}} (|\bm{r}_2 - \bm{R}| ) \, 
           \left\{ 1 - \frac{\bm{r}_2 \cdot \bm{R}}{R^2} \right\} \, \rho _n^{\textrm{NF}} (r_2) \, d \bm{r}_2 \;,
\label{eq:u6bfls}
\end{eqnarray}
where $\rho_n^{\textrm{NF}}$ and $\rho_p^{\textrm{NF}}$ denote point neutron and proton
densities, respectively.

\subsection{Numerical evaluation of $p$-$^6$He potentials} \label{sc:num}


To evaluate the $p$-$^6$He folding potentials as specified in the preceding
section, we have to fix the following elements; the $p$-$\alpha$
interaction $V_{p\alpha}$, the $p$-$p$ and $p$-$n$ interactions $V_{pp}$ and
$V_{pn}$, and the densities in $^6$He, $\rho_{\alpha}$, $\rho_p$ and
$\rho_n$.
These are discussed in the following subsections, respectively.


\subsubsection{$p$-$\alpha$ interactions}

\begin{figure}[tb]
\includegraphics[width=0.9\linewidth,clip]{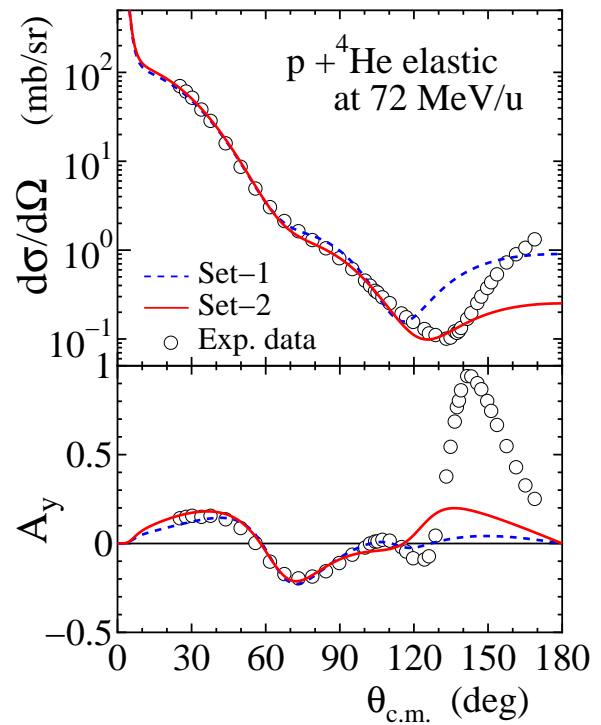}
\caption{(Color online) Angular distribution of the cross section and $A_y$ 
 for the $p+^4$He elastic scattering at 72~MeV/nucleon.
 The solid and dashed lines are the optical model calculations
 with Set-1 and Set-2 parameters, respectively.
 The experimental data are taken from Ref.~\cite{Burzynski89}.}
\label{fig:pHe4Xs}
 \end{figure}

\begin{figure}[tb]
\includegraphics[width=0.75\linewidth,clip]{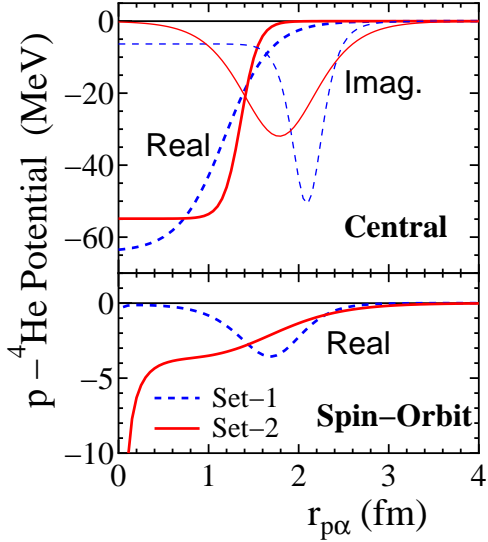}
\caption{(Color online) Optical potentials for the $p$+$^4$He elastic scattering at 72~MeV/nucleon.
  The thick dashed (solid) lines are for the real part of Set-1 (Set-2) potential
  and the thin dashed (solid) lines are for the imaginary part.
}
\label{fig:pHe4Pot}
 \end{figure}

For $V_{p \alpha}$ used in the CF potential, we assume the standard WS
potential such as given in Eq.~(\ref{eq:op1}).  
The parameters involved are searched so as to fit the data 
of $d \sigma / d \Omega$ 
and $A_y$
in the $p$+$\alpha$ scattering at 72~MeV/nucleon \cite{Burzynski89}.
Particular attention was given to reproducing the observables
in the forward angular region, since overall agreements with the data are 
not found
in spite of the careful search of the parameters.
Two typical parameter
sets, with and without the volume absorption term, are labeled as Set-1
and Set-2 in Table~~\ref{tab:pHe4Pot}.
The results of calculations made with these potentials are compared with 
the data in
Fig.~\ref{fig:pHe4Xs}, where the solid and dashed lines show those by Set-1
and Set-2 potentials, respectively.
Both calculations describe the data up to $\theta \approx 100^\circ$
but do not reproduce those at backward angles, $\theta \agt 120^\circ$.
Such discrepancies between the calculated results and the measured 
data at the backward angles
suggest participation of contributions of other reaction mechanisms,
such as knock-on type exchange scattering of the proton with target nucleons.
Such possible extra mechanisms will be disregarded at present
since we are concerned with the $p$-$\alpha$ one-body potential.
In our CF calculations, we adopt the potentials with the above parameter sets
as $V_{p \alpha}$.
However, the validity of the CF potential thus obtained is limited to forward
scattering  angles,
a low momentum transfer region, of $p$+$^6$He scattering.
The real and imaginary parts of $V_{p \alpha}^0$ and the real part of 
$V_{p \alpha}^{\textrm{LS}}$
for the above parameter sets are displayed in the upper and lower panels
of Fig.~\ref{fig:pHe4Pot}.
Although Set-1 (dashed) and Set-2 (solid) potentials have rather different
$r_{p\alpha}$ dependence, as shown later, this difference is moderated
in the folding procedure so yielding similar CF potentials.

\begin{table*}[bthp]
\caption{Parameters for the optical potentials for $p$+${}^4$He
at 72~MeV/nucleon.}
\begin{ruledtabular}
\begin{tabular}{c|ccccccccccccc}
    &  $V_0$  &  $r_{0r}$  &  $a_{r}$  &  $W_0$  &  $r_{0i}$ &  $a_{i}$
    &  $W_d$  &  $r_{0id}$ &  $a_{id}$
    & $r_{0\textrm{C}}$ & $V_{s}$ & $r_{0s}$ & $a_{s}$ \\
    &  (MeV)  &  (fm)  &  (fm)  &  (MeV)  &  (fm)  &  (fm)
    &  (MeV)  &  (fm)  &  (fm)
    & (fm) &  (MeV)  &  (fm)  &  (fm) \\ \hline
Set-1 & 64.13 & 0.7440 & 0.2562 & 6.338 & 1.450 & 0.2089
      & 46.23 & 1.320  & 0.1100 & 1.400 & 2.752 & 1.100 & 0.2252 \\
Set-2 & 54.87 & 0.8566 & 0.09600 & ---  & ---   & --- 
      & 31.97 & 1.125  & 0.2811 & 1.400 & 3.925 & 0.8563 & 0.4914\\
\end{tabular}
\end{ruledtabular}
\label{tab:pHe4Pot}
\end{table*}

\subsubsection{$p$-$p$ and $p$-$n$ interactions}

For $V_{pn}$ and $V_{pp}$ used in the CF and NF potentials,
we adopt the complex effective interaction, CEG~\cite{Yamaguchi83,Nagata85,Yamaguchi86},
where the nuclear force~\cite{Tamagaki68} is modified by the medium effect
which takes account of the virtual excitation of nucleons of the nuclear matter
up to $10 k_\textrm{F}$ by the $g$-matrix theory.
The nuclear force is composed of Gaussian form factors and the parameters contained
are adjusted to simulate the matrix elements of the Hamada-Johnston potential~\cite{Hamada62}.
The CEG interaction has been successful in reproducing $d\sigma / d\Omega$ and $A_y$
measured for the proton elastic scattering by many nuclei
in a wide incident energy range, $E_p=$ 20--200~MeV,
in the framework of 
the folding model~\cite{Yamaguchi83,Nagata85,Yamaguchi86}.
It has been shown that the imaginary part of the folding
potential given by the CEG interaction is slightly too large
to reproduce experimental N-A
scattering~\cite{Yamaguchi83,Yamaguchi86}.
In the present calculation, therefore,
we adopt the normalizing factor $N_I=0.7$
for the imaginary part of the CEG interaction.
However, calculations with $N_I=1.0$ do not give an essential change
to the results.

\subsubsection{Densities of $\alpha$, $p$ and $n$ in $^6$He}

\begin{figure}[tb]
\includegraphics[width=0.80\linewidth,clip]{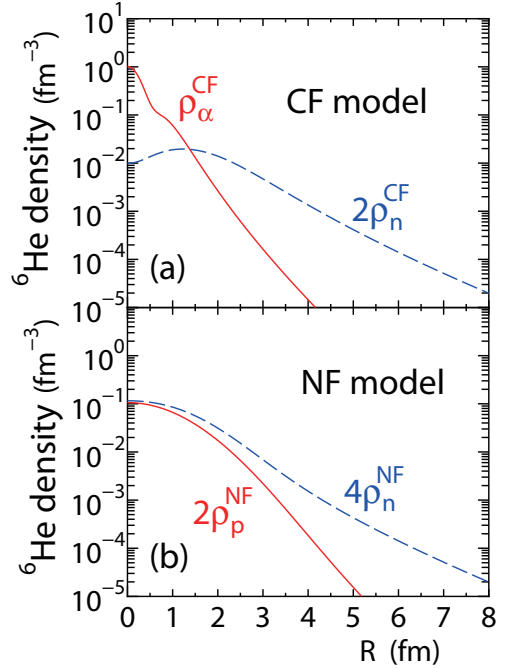}
\caption{(Color online) The densities of $\alpha$, $p$, and $n$ in $^6$He used in folding models.
  The $\rho_\alpha^{\textrm{CF}}$ and $\rho_n^{\textrm{CF}}$ in panel \textbf{(a)} are used in the CF calculation.
  The $\rho_p^{\textrm{NF}}$ and $\rho_n^{\textrm{NF}}$ in panel \textbf{(b)} are used in the NF calculation.
  All densities are normalized as
  $4 \pi \, \int \rho_\textrm{x} \, r^2 \, dr = 1$,
  where $\textrm{x} = \alpha$, $n$, and $p$.
 }
\label{fig:He4Dens}
 \end{figure}

The densities $\rho _n^{\textrm{CF}}$ and $\rho_\alpha^{\textrm{CF}}$
for the CF calculation are obtained by applying the Gaussian expansion
method~\cite{Hiyama96,Hiyama03} to the $\alpha n n$ cluster model of $^6$He.
This method has been successful in describing structures
of various few-body systems as well as $^6$He~\cite{Hiyama96,Hiyama03}.
As for the $n$-$n$ interaction, we choose AV8' interaction~\cite{Wiringa84}.
It is reasonable to use a
bare (free space) $n$-$n$ interaction between
the two valence neutrons in $^6$He as they are dominantly
in a region of low density.
As for the $\alpha$-$n$ interaction, we employ
the effective $\alpha$-$n$ potential in Ref.~\cite{Kanada79},
which was designed to reproduce well the low-lying states
and low-energy-scattering phase shifts of the $\alpha$-$n$ system.
The depth of the $\alpha$-$n$ potential is modified slightly
to adjust the ground-state binding energy of $^6$He to the empirical value.
In Fig.~\ref{fig:He4Dens}(a), the densities obtained are shown  as functions of $r$,
the distance from the center of mass of $^6$He,
where $\rho _\alpha^{\textrm{CF}}$ is localized in a relatively narrow region around the center,
while $\rho _n^{\textrm{CF}}$ is spread widely.

The NF calculation depends on the assumptions made for the densities of the 
two protons and four neutrons in $^6$He as well as those made for
the $p$-$p$ and $p$-$n$ interactions~\cite{Uesaka10}.
At present, to see the essential role of clustering the four nucleons into 
the $\alpha$-particle core,
we use the densities of the proton and the neutron in the 
$\alpha$ obtained by decomposing the density of the point
$\alpha$, $\rho _\alpha^{\textrm{CF}}$, to the densities of the
constituent nucleons with a one-range Gaussian form factor with range
1.40~fm.
The total nucleon densities of $^6$He, $\rho _p^{\textrm{NF}}$, and 
$\rho _n^{\textrm{NF}}$,
where the latter includes the contribution of the valence neutrons, 
are displayed in Fig.~\ref{fig:He4Dens}(b).
The neutron density $\rho _n^{\textrm{NF}}$ has a longer tail than the 
proton one $\rho _p^{\textrm{NF}}$
due to the presence of the valence neutrons.
In Refs.~\cite{ItagakiPrv,Itagaki03} the nucleon densities of $^6$He 
were calculated in a more sophisticated way. They
produced  densities similar to the present ones for the protons and
neutrons. 
These two kinds of nucleon densities
provide similar results in the NF calculation of
$d\sigma / d\Omega$ and $A_y$ of the $p+^6$He scattering.
Thus, in the following, we will discuss $U_{\textrm{NF}}$
as formed using the densities shown in Fig.~\ref{fig:He4Dens}(b).

\subsubsection{$p$-$^6$He folding potentials}
\begin{figure*}[htb]
\includegraphics[width=0.85\linewidth,clip]{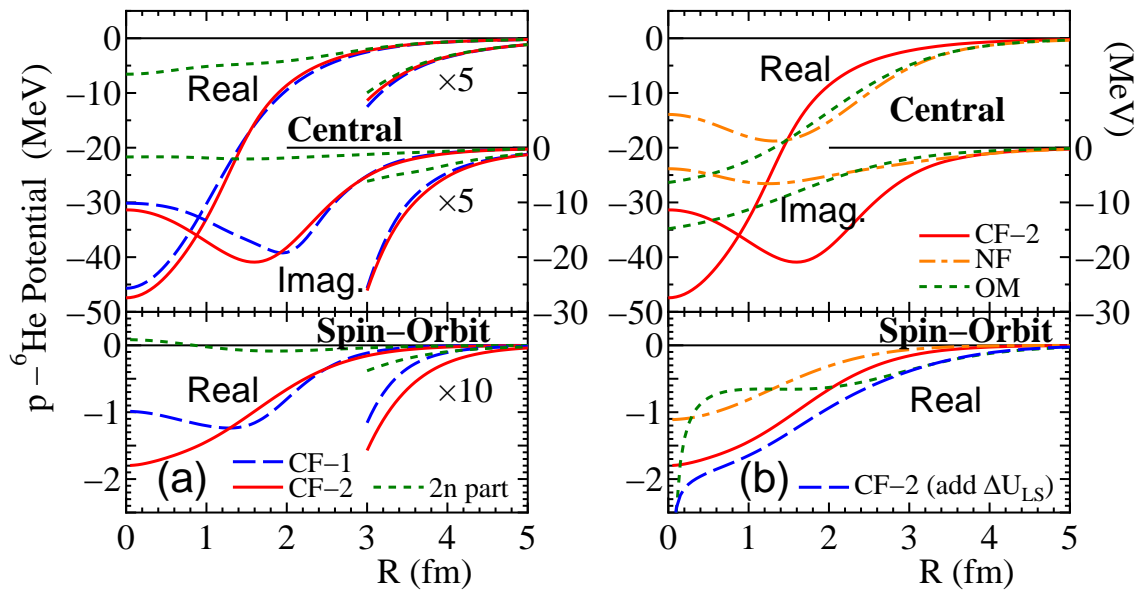}
\caption{(Color online) Potentials between proton and $^6$He.
 \textbf{(a)}: The long-dashed (solid) lines are the CF calculation with Set-1 (Set-2) parameters
 for $V_{p \alpha}$.
   The short-dashed lines show the contribution of the valence two
  neutrons to the CF potential.
 \textbf{(b)}: The solid lines are the CF calculation with Set-2 parameters,
   the dot-dashed lines are the NF one and the short-dashed lines are the phenomenological
   optical potential.
   The CF spin-orbit potential corrected by $\Delta U_{\textrm{LS}}$ term
   is shown by the long-dashed line (see text for detail).
}
\label{fig:pHe6Pot}
 \end{figure*}

In Fig.~\ref{fig:pHe6Pot}, the resultant $p$-$^6$He potentials,
$U_{\textrm{CF}}$ and $U_{\textrm{NF}}$, are compared with each other
as well as with
the optical model potential $U_{\textrm{OM}}$.
The CF potentials calculated by the two sets of $V_{p \alpha}$ in Table~\ref{tab:pHe4Pot},
say $U_{\textrm{CF-1}}$ and $U_{\textrm{CF-2}}$, 
are shown by long-dashed and solid lines in 
Fig.~\ref{fig:pHe6Pot}(a), respectively.
The folding procedure gives similar results in both cases.
The contribution of $V_{pn}$ is displayed by short-dashed lines in the
figure, which is found to be mostly small.
Especially, in the spin-orbit potential, the contribution from $V_{pn}$ 
is one order of magnitude smaller than that from $V_{p\alpha}$.
The main contribution to $U_{\textrm{CF}}$
arises from the interaction $V_{p\alpha}$ except for the 
central real potential at $R\agt$3~fm, which is 
dominated by the $V_{pn}$ contribution.
This is supposed to be the reflection of the extended neutron density shown in
Fig.~\ref{fig:He4Dens}
and produce significant contributions to the observables as discussed
later.

In Fig.~\ref{fig:pHe6Pot}(b), $U_{\textrm{CF}}$ due to Set-2 of $V_{p \alpha}$,
$U_{\textrm{NF}}$, and $U_{\textrm{OM}}$ are shown by solid, dot-dashed,
and short-dashed lines, respectively.
First we consider the central part of the potentials.
For small $R$, the real part of $U_0^{\textrm{CF}}$ is deeper than
those of $U_0^{\textrm{NF}}$ and $U_0^{\textrm{OM}}$,
while for large $R$, $U_0^{\textrm{CF}}$ is shallower than other two.
In the imaginary part, the magnitude of $U_0^{\textrm{CF}}$ is much bigger
than those of the other two potentials.
This will compensate the deficiency of the real part of $U_0^{\textrm{CF}}$ at large $R$,
for example in the calculation of the cross section.
On the other hand, the spin-orbit part of $U_{\textrm{OM}}$ has larger magnitude
for $R \agt 2$ fm and thus has a longer range compared with those of other two potentials.
Such long-range nature of the spin-orbit interaction is a characteristic
feature of the spin-orbit part of $U_{\textrm{OM}}$ as described in
Section~\ref{sc:phen}.
This is discussed later in more detail with relation to $A_y$.

\subsection{Comparison between experiments and calculations in $\bm{p}$+$^6$He scattering}

\begin{figure*}[tbp]
\includegraphics[width=0.4\linewidth,clip]{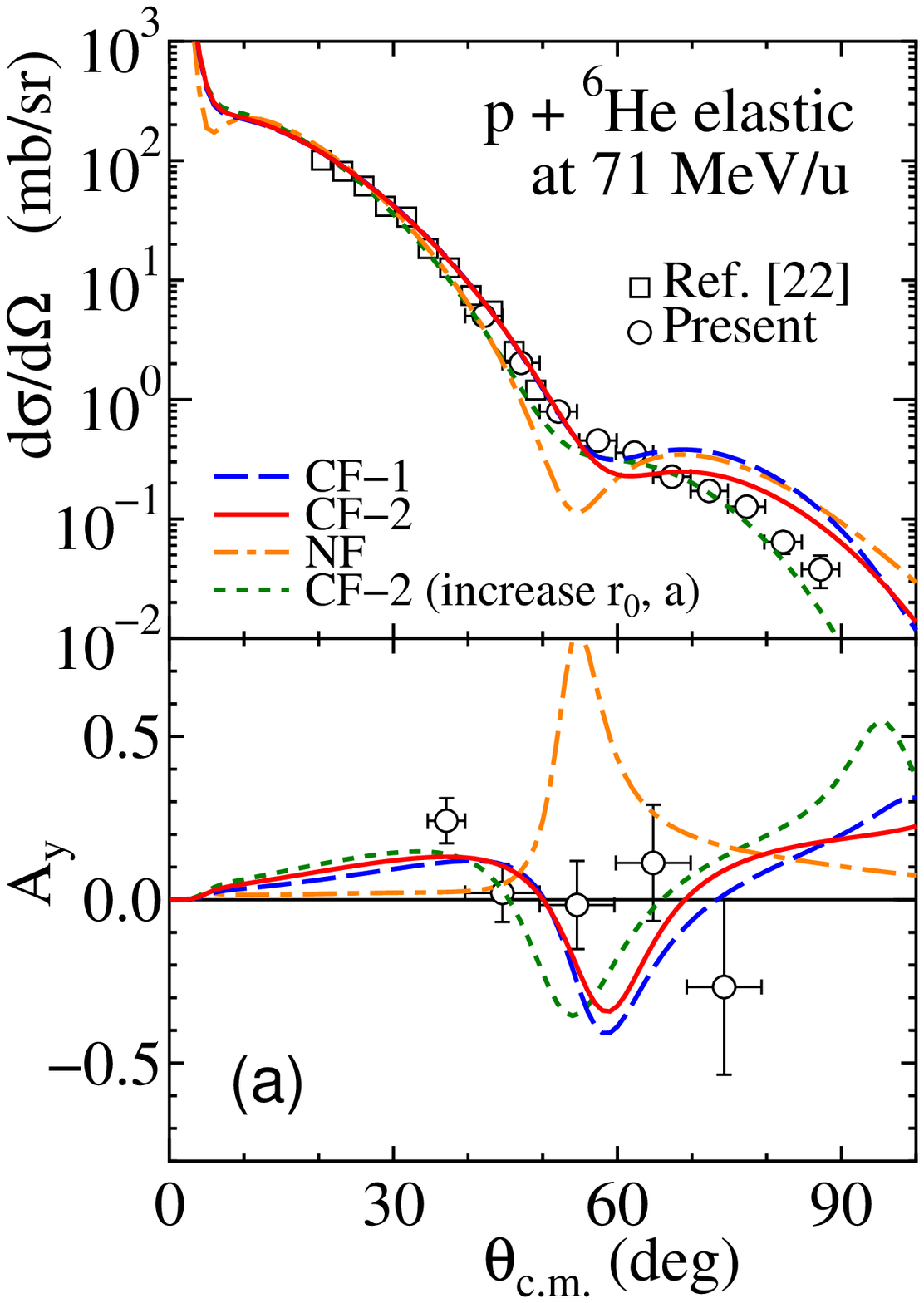}
\includegraphics[width=0.4\linewidth,clip]{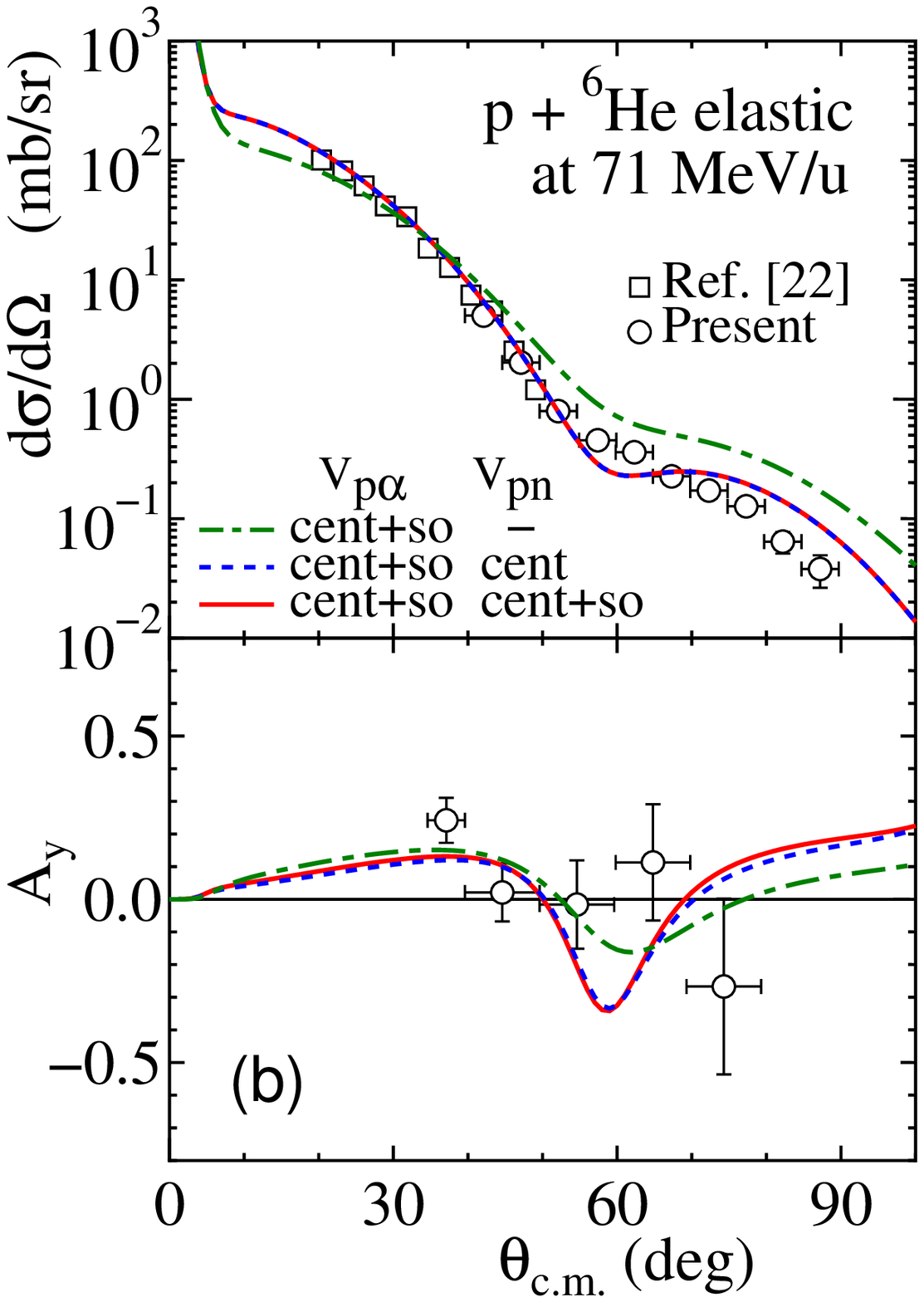}
\includegraphics[width=0.4\linewidth,clip]{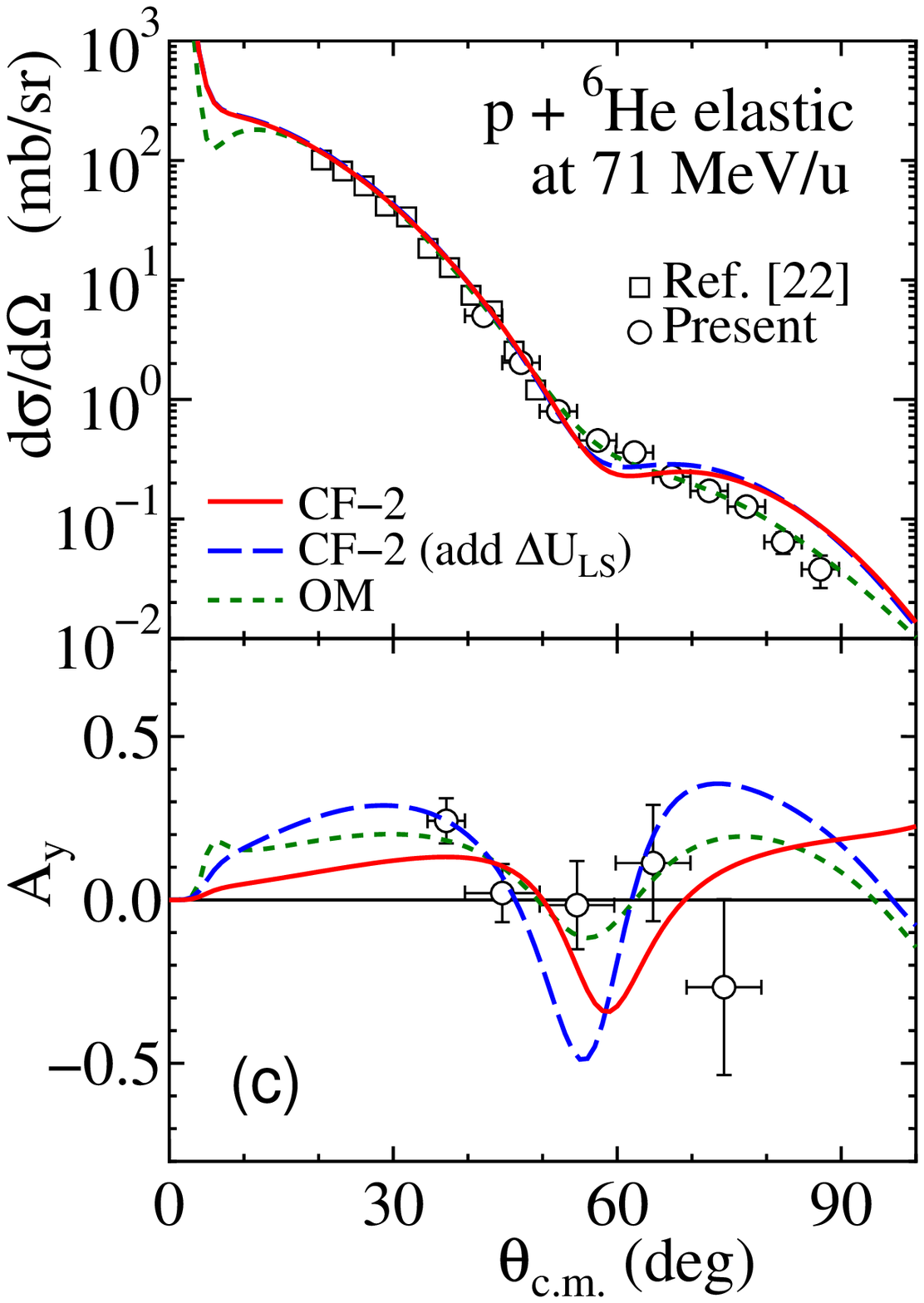}
\hspace*{0.4\linewidth}
\caption{(Color online) Angular distribution of the cross section and $A_y$ 
 for the $p$+$^6$He elastic scattering at 71~MeV/nucleon.
 The experimental data are denoted by circles (present) and squares
  (Ref.~\cite{Korsheninnikov97}).
 \textbf{(a)}: The long-dashed (solid) lines are the CF calculation
   in which Set-1 (Set-2) parameters are used for $V_{p \alpha}$.
   The dot-dashed lines are the NF calculation.
 \textbf{(b)}: 
The dashed lines and the solid ones include the $V_{pn}$ interaction,
  where the formers neglect the spin-orbit part of $V_{pn}$.
The dash-dotted lines include only $V_{p\alpha}$ interaction.
 \textbf{(c)}: The solid lines are the CF calculation with Set-2 parameters for $V_{p \alpha}$
   and the short-dashed lines are the OM calculation.
The long-dashed lines are the CF calculation with Set-2 parameters for
  $V_{p\alpha}$ where $\Delta U_{\textrm{LS}}$ is added (see text for
  detail).}
\label{fig:pHe6Xs}
 \end{figure*}

In the following, the $d\sigma / d\Omega$ and $A_y$ 
for $p$-${}^6$He elastic scattering
calculated  using 
$U_{\textrm{CF}}$, $U_{\textrm{NF}}$, and $U_{\textrm{OM}}$ are compared 
with the data taken at 71~MeV/nucleon.
In Fig.~\ref{fig:pHe6Xs}(a) the results obtained using the two CF potentials,
$U_{\textrm{CF-1}}$ and $U_{\textrm{CF-2}}$, 
are shown by long-dashed and solid lines, respectively.
Both results are very similar to each other and well describe the data of
$d\sigma / d\Omega$,
except for large angles where the calculations overestimates the data by
small amounts.
The calculations also describe the angular dependence of the measured 
$A_y$ up to $\theta \simeq 55^\circ$.
These successes basically support the CF potential as a reasonable
description of the scattering.
The discrepancies at large angles,  i.e. a large momentum transfer
region, 
may be related to the limitation of the validity of $V_{p \alpha}$ used
in the folding, as discussed in the subsection~\ref{sc:num}.

In Fig.~\ref{fig:pHe6Xs}(a),
the results of the calculation made using the NF potential 
are shown by dot-dashed lines. They do not
reproduce the data well.
The calculation gives an deep valley around $\theta \simeq 53^\circ$ 
in the angular
distribution of $d\sigma / d\Omega$ and 
a large positive peak at the corresponding angle of the $A_y$ angular
distribution.
These features do not exist in the data.
Since the present nucleon densities originated from the CF model
ones, the essential difference between the CF and NF potentials will be
produced by the use of the different interactions.
Thus, the CF calculation will owe its successes to the inclusion of the
characteristics of the realistic $p$-$\alpha$ interaction into the
$p$-$^6$He potential.

It is interesting to examine if the $\alpha$ core in $^6$He
is somewhat diffused compared with a free $\alpha$-particle,
due to the interactions from the valence neutrons.
For that purpose, we increased the radius and  diffuseness parameters,
$r_0$ and $a$, in $V_{p \alpha}$ potential as $r_0$ to $1.1 \, r_0$ and
$a$ to $a+0.1$ fm. The depth parameters were changed to keep constant
the values of the corresponding volume integrals.
The effect of this change is shown by the short-dashed lines in
Fig.~\ref{fig:pHe6Xs}(a), where 
reproduction of the data is improved somewhat, especially in
$d\sigma/d\Omega$ at large angles.

In Fig.~\ref{fig:pHe6Xs}(b), the contributions of the valence neutrons
are demonstrated for the CF-2 calculation.
As is speculated from the analyses of the form factors of the potential in
Fig.~\ref{fig:pHe6Pot}(a), the dominant contribution to the observables
in the CF calculation arises from $V_{p\alpha}$ displayed by the
dot-dashed lines in Fig.~\ref{fig:pHe6Xs}(b).
However, the valence neutrons 
produce indispensable corrections to the observables.
That is, the $pn$ central interaction decreases $d\sigma/d\Omega$ at large
angles, giving remarkable improvements of the agreement with the data as
shown by the dashed lines.
The $pn$ interaction also contributes to $A_y$ by a considerable amount
through the central part.
 A detailed examination of the calculation revealed that such corrections
were due to the $V_{pn}$ part of the folding central potential in
a $R$ region between $R=$2~fm and 4~fm (see
Fig.~\ref{fig:pHe6Pot}(a)).
The spin-orbit part of $V_{pn}$ gives almost no effect to the
observables as shown by the solid lines in Fig.~\ref{fig:pHe6Xs}(b).
This is consistent with the result shown by Crespo {\it et
al.}~\cite{Crespo07} in a study at higher incident energy.

In Fig.~\ref{fig:pHe6Xs}(c), we compared the results of the CF-2 
calculation (solid
lines) with those of the
OM calculation in the preceding section (long-dashed lines)
as well as with the data.
In the optical model analysis, the experimental data can be reproduced
only with a shallow and long-ranged spin-orbit potential.
Compared with this potential, the spin-orbit part of CF-2 potential 
has a shorter range as displayed by a solid line in the lower panel of
Fig.~\ref{fig:pHe6Pot}(b).
To investigate the role of the long tail in the spin-orbit
potential, we calculated the observables by 
adding a weak but long-range spin-orbit interaction $\Delta
U_{\textrm{LS}}$ to the CF interaction.
This correction is assumed to be the Thomas type
as 
\begin{equation}
\Delta U_{\textrm{LS}}(R) = v_\textrm{add} \, \frac{2}{R} \, 
    \frac{d}{dR} \, 
    \left[ 1 +  \exp \left\{ (R - 6^{1/3} r_\textrm{add} ) / a_\textrm{add} \right\} \right] ^{-1} \;.
\label{eq:delULS}
\end{equation}
For simplicity, we adopt $v_\textrm{add} = 1$~MeV, $r_\textrm{add} = 1.5$ fm, and
$a_\textrm{add} = 0.7$ fm where the large magnitudes of $r_\textrm{add}$
and $a_\textrm{add}$ are consistent with the characteristics of the
magnitudes of $r_{0s}$ and $a_s$ of the OM potential discussed in
Sec.~\ref{sc:phen}.
The calculated observables are displayed in Fig.~\ref{fig:pHe6Xs}(c) by
short-dashed lines, 
where $d\sigma / d\Omega$ is little affected but $A_y$ receives a drastic change,
i.e. the angular distribution of $A_y$ is now similar to that by the OM calculation
in a global sense
showing
qualitative improvements in comparison with the data.
To see the contribution of $\Delta U_{\textrm{LS}}$ to the potential,
we plot $U_{\textrm{LS}}^{\textrm{CF}} + \Delta U_{\textrm{LS}}$ in
Fig.~\ref{fig:pHe6Pot}(b) by long-dashed lines, 
where the new spin-orbit potential becomes very close to that of the OM potential
at $R \agt 2.5$ fm.
It is indicated that the long tail of the spin-orbit potential is particularly
important in reproducing the angular distribution of $A_y$, while its
microscopic origin is still to be investigated.
When some corrections which increase the 
range of the spin-orbit interaction are found, they will be 
effective for improving the CF calculation.


\section{Microscopic Model Analyses}

In this section, we will describe the theoretical analysis of the
present data by a microscopic model developed in Ref.~\cite{Amos00}.
In this model, one can predict the scattering observables
such as cross sections and analyzing powers with one run of the relevant
code (DWBA98) with no adjustable parameter. 
Complete details as
well as  many examples of  use of this coordinate  space microscopic
model approach are to be  found in the review~\cite{Amos00}. Use of
the complex,  non-local, nucleon-nucleus optical  potentials defined in
that  way,  without localization  of  the  exchange amplitudes, has
given 
predictions of  differential cross sections and  spin observables that
are in good agreement with data from many nuclei ($^3$He to $^{238}$U)
and for  a wide  range of  energies (40 to  300~MeV). Crucial  to that
success is  the use  of effective nucleon-nucleon  ($NN$) interactions
built upon  $NN$ $g$-matrices.
The
effective $NN$ interactions are complex, energy and density dependent,
admixtures  of Yukawa  functions.
They  have central,  two-nucleon
tensor  and   two-nucleon  spin-orbit  character.   The  $NA$  optical
potentials result  from folding those effective  interactions with the
one-body density matrix elements (OBDME) of the ground state in
the  target nucleus.  Antisymmetrization  of the  projectile with  all
target nucleons  leads to  exchange amplitudes, making
the microscopic  optical potential non-local.  
For brevity,  the optical  potentials that  result are  called $g$-folding
potentials.
Another application has been in the prediction of integral
observables
of elastic scattering of both protons and neutrons, with equal success
\cite{De01}. Thus, the method is  known now to give good predictions of
both angular-dependent  and  integral  observables. 

It is important to note that the level of agreement with data in the
$g$-folding approach  depends on  the quality  of  the structure 
model that is used.
Due to the character  of
the hadron
force, proton  scattering is  preferentially sensitive to  the neutron
matter  distributions  of  nuclei;  a  sensitivity seen  in  a  recent
assessment,    using   proton    elastic   scattering,    of   diverse
Skyrme-Hartree-Fock model  structures for $^{208}$Pb~\cite{Ka02}. 

\subsection{Structure of ${}^6$He used}

$^6$He is a  two-neutron halo nucleus and  has been described  well by  shell  
model calculations.  In calculation of the $g$-folding potential for 
protons interacting with ${}^6$He, a  complete $(0$+$2$+$4)\hbar\omega$ 
shell model calculation has been made to specify
the ground state OBDME. 
Essentially they are the  occupation numbers which define the matter 
densities of the nucleus.

\begin{figure}[htbp]
 \begin{center}
  \includegraphics[width=0.775\linewidth]{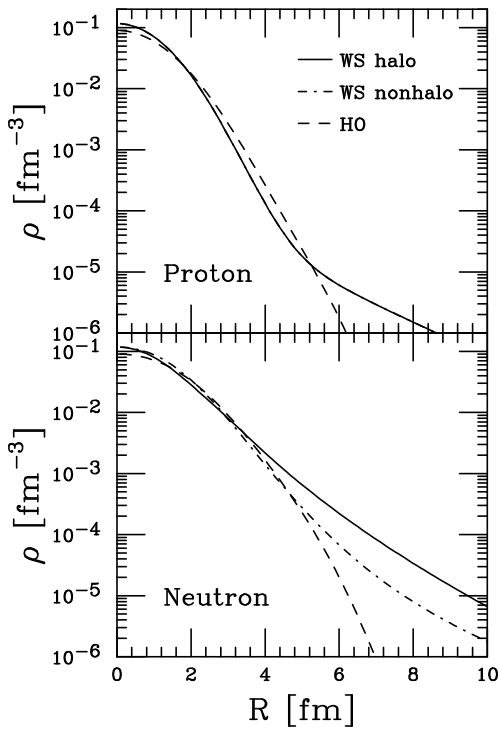}
  \caption{The (model) proton and neutron densities are shown in the
  upper and lower panels, respectively.
  The solid, dot-dashed, and dashed curves represent those obtained with
  WS halo, WS nonhalo, and HO sets, repectively.}
  \label{density}
 \end{center}
\end{figure}

In the present study, we assume three sets of the single-nucleon (SN)
wave functions for $^6$He.
One is the oscillator wave functions with an oscillator length of
2.0~fm (HO set).
However, a neutron-halo character of $^6$He can not be given by the
oscillator wave function whatever oscillator length is used as shown by
the dashed curve in the lower panel of Fig.~\ref{density}.
Thus, we assume two sets of SN wave functions defined in Woods-Saxon
(WS) potentials.
One of them is obtained by taking the geometry of the potential from
that found appropriate in Ref.~\cite{Amos00},
where electron form factors and proton scattering from ${}^{6,7}$Li are
studied.
That study provided a set of SN wave functions that we specify as 
WS nonhalo set since the $p$-shell nucleons were all reasonably bound.
The extended neutron matter character of ${}^6$He is found by 
choosing the binding energy of  the halo-neutron orbits to give
the single-neutron separation energy  (1.8~MeV) to the  lowest energy
resonance in $^5$He. The set of SN wave functions that result are
specified as WS halo set. The  associated density profile has the extensive
neutron density coming from the halo. 
Density profiles given by the various sets of SN wave functions are 
shown in Fig.~\ref{density}. 
The dashed, solid, and dot-dashed curves show the density
distributions of HO, WS halo, and WS nonhalo sets, respectively.
The difference between proton distributions of WS halo and WS nonhalo
sets can not be seen.

Use of WS halo set in analyses of 40.9~MeV/nucleon data~\cite{Lagoyannis01} gave a
value of 406 mb for the reaction cross section, which is in good
agreement with the measured value.
Additional evidence for WS halo set is given by 
the root  mean   square  (r.m.s.)  radius of the matter distribution,
which is most  
sensitive  to characteristics of the outer surface of a nucleus.  Using 
WS nonhalo set of the SN wave functions gave an r.m.s. radius  
for ${}^6$He of 2.30~fm, which is much smaller than the expected value
of 2.54~fm.
On the other hand, using WS halo set gave 
an r.m.s. radius for ${}^6$He of 2.59~fm in good
agreement with that expectation.

\subsection{Differential cross sections and analyzing powers}

\begin{figure}[htbp]
 \begin{center}
  \includegraphics[width=0.87\linewidth]{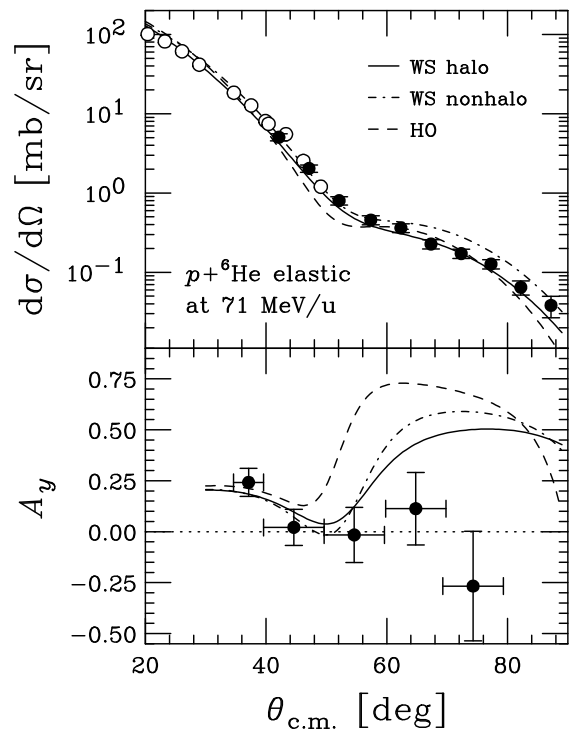}
  \caption{Differential cross sections and analyzing
  powers of the 
  $p+^6$He elastic scattering at 71~MeV/nucleon (open circles:
  Ref.~\cite{Korsheninnikov97}, closed circles: present work).
  Three curves are results of $g$-matrix folding calculation with
  $^6$He densities presented in Fig.~\ref{density}.
}
  \label{fig:miccalc}
 \end{center}
\end{figure}

The  cross sections and  analyzing powers for 
the $p+^6$He elastic scattering  at 71~MeV/nucleon are shown in the top and
bottom panels of
Fig.~\ref{fig:miccalc}, respectively. The calculated results shown 
therein by the dashed lines were found using the $g$-folding  potential
obtained with HO set of SN wave functions.
This calculation does not give a
satisfactory result; especially in the case of the analyzing power.
The solid curves show results found using WS halo set while those
depicted by the dot-dashed curves are those found with WS nonhalo set.
Of these the halo description gives the better match to data  
especially at the larger scattering angles.
This result is consistent with the findings from
analyses of lower energy scattering data at 40.9~MeV/nucleon~\cite{Lagoyannis01}
and at 24.5~MeV/nucleon~\cite{Stepantsov02}.

In fact, the differential cross sections calculated with WS halo set
match the data so well that one does not need to contemplate any
adjustment.
However, the story is not so simple when one also considers the
analyzing power data.
At forward scattering  angles, both 
WS sets reasonably match the data. But neither WS result produces the 
distinctive trend of small values found at larger 
scattering angles. Nonetheless the best result is that found on using
WS halo set of SN functions.  Given that the cross section values in the 
region of 60$^\circ$ to 90$^\circ$ is of an order of 0.1 mb/sr, the limitations
in the present microscopic model formulation of the reaction dynamics 
may be the problem.


\section{Summary}

The vector analyzing power has been measured for the elastic scattering of
$^6$He from polarized protons at 71~MeV/nucleon to investigate the 
characteristics of the spin-orbit potential between the proton and the
$^6$He nucleus.
Measurement of the polarization observable was realized 
in the RI-beam experiment by using the newly constructed solid
polarized proton target, which can be operated in a low magnetic field
of 0.1~T
and at high temperature of 100~K.
The measured $d\sigma/d\Omega$ of the $p+^6$He elastic scattering were
almost identical to those of the $p+^6$Li.
On the other hand, the $A_y$ were found to be largely different from
those of the $p+^6$Li and rather similar to those of the $p+^4$He
elastic scattering.

To extract the gross feature of the spin-orbit interaction
between a proton and $^6$He, an optical model potential was
determined phenomenologically by fitting the experimental data of
$d\sigma/d\Omega$ and $A_y$.
Compared with the global systematics of the potentials for stable
nuclei, it is indicated that the spin-orbit potential for $^6$He is 
characterized by a small value of $V_s$ and large values of $r_{0s}$ and
$a_s$, namely by a shallow and long-ranged radial shape.
Such characteristics might be the reflection of the diffused density of
the neutron-rich $^6$He nucleus.

The cluster folding calculation was carried out to 
get a deeper insight into the optical potential, assuming the 
$\alpha$+$n$+$n$ cluster structure for $^6$He.
In addition, nucleon
folding calculations were also performed 
by decomposing the $\alpha$ core into four nucleons.
The experimental data could not be reproduced by the nucleon
folding calculation, whereas the $\alpha n n$ cluster folding calculation
gives the reasonable agreements with the data.
Thus, this indicates that it is important to take into account of the
$\alpha$-clusterization in the description of $p+^6$He elastic
scattering.
The cluster folding calculation shows that the dominant contribution
to the $p$-$^6$He potential arises from the interaction between the proton and
the $\alpha$ core.
Especially, in the spin-orbit potential, the contribution of the
interaction between the proton and valence neutrons was found to be much
smaller than the $\alpha$ core contribution.
However, the measured cross section at large angles can not be
understood without the contribution from the scattering by the valence
neutrons.
Comparison of the phenomenological optical potential and the cluster folding
one indicates that the long-range nature of the spin-orbit potential is
important in reproducing the $A_y$ data at large angles.
The microscopic origin of such a long tail is still to be investigated.

The data were also compared with the predictions obtained from a fully
microscopic $g$-folding model.
Three sets of single nucleon wave functions were tried since other
details of 
the calculation were predetermined.
The model, which has been successful in analyzing $p$+$^6$He
scattering cross sections in the past~\cite{Ka02}, again gives 
good reproduction of the data 
in the present case when the bound state wave functions 
specify that ${}^6$He has a neutron halo.
However, the match to the data, in particular the analyzing power, is
not perfect.
This may indicate limitation of the structure model used and/or of
unaccounted reaction mechanisms that influence the larger momentum
transfer results.

This work has demonstrated the capability of the solid polarized proton
target in low magnetic field and high temperature to probe the new
aspects of the reaction involving unstable nuclei.
Future polarization studies of such kinds will provide us with valuable
information on the reaction and structure of unstable nuclei.

\section*{Acknowledgments}

We thank the staffs of RIKEN Nishina Center and CNS for
the operation of the accelerators and ion source during the measurement.
S.~S. acknowledges financial support by a Grant-in-Aid for JSPS Fellows
(No.~18-11398).
This work was supported by the Grant-in-Aid for Scientific Research
No.~17684005 of the Ministry of Education, Culture, Sports, Science, and
Technology of Japan.

\appendix

\section{Absolute measurement of target polarization} \label{sec:calib}

In the case of conventional solid polarized targets, the NMR signal 
usually is related to
the absolute magnitude of the polarization by measuring the target
polarization under the state of thermal equilibrium (TE).
However, measurement of the TE polarization is quite difficult in 
our target.
The first reason for this is that the TE polarization is very small 
in a low magnetic field and at high temperature, since it is represented
by $P_{\textrm{TE}}=\tanh \left( \frac{\mu B}{2kT} \right)$, where $\mu,
B$, and $T$ are the magnetic moment of proton, the field strength, and
the temperature, respectively.
The second reason is that the sensitivity of the present NMR system is
not sufficiently high since the target design is optimized for
scattering experiments.

One of the simple methods to measure the absolute target polarization
would be the measurement of the spin-dependent asymmetry $\epsilon = P_y
A_y$ for the proton elastic scattering whose analyzing power $A_y$ are 
known.
In the present study, we measured the spin-asymmetry for the $p+^4$He elastic
scattering at 80~MeV/nucleon.
The $A_y$ have already been measured by Togawa {\it et
al}~\cite{Togawa87}.
The use of the $p+^4$He scattering is profitable since we can measure
the $\epsilon$ 
with the same experimental setup as that for the $p+^6$He measurement
only by changing settings of the fragment separator RIPS 
to produce a secondary $^4$He beam.
The profile of the $^4$He beam on the target was tuned to be almost same as
that of the $^6$He beam.

\begin{figure}[htbp]
 \begin{center}
  \includegraphics[width=0.82\linewidth]{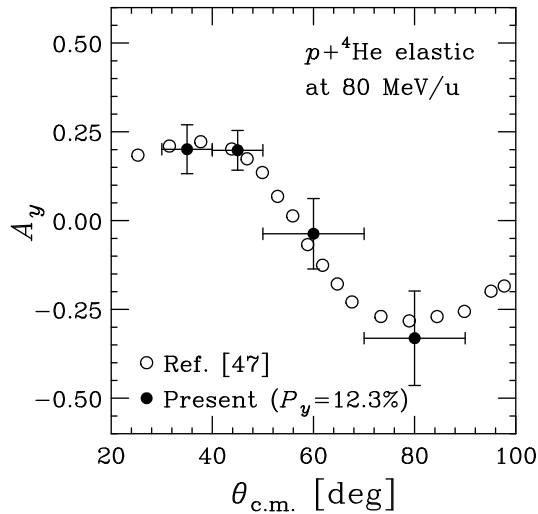}
  \caption{
  Analyzing powers of the $p+^4$He elastic scattering
at 80~MeV/nucleon. Closed circles indicate the present data where $P_y=12.3\%$
  is assumed. Open circles represent the reference data taken from 
  Ref.~\cite{Togawa87}.
  }
  \label{fig:calib}
 \end{center}
\end{figure}

Figure~\ref{fig:calib} shows $A_y$ of the $p+^4$He elastic scattering
at 80~MeV/nucleon.
The open circles represent the previous data~\cite{Togawa87}, while the
closed ones
show the present data whose magnitudes are scaled to the previous ones.
From the scaling factor, 
the average polarization during the $p+^4$He measurement was determined to
be $P_y=12.3\pm2.4$\%.
The relative uncertainty of the polarization $\Delta P_y / P_y$, 
which was 19\% in the present work, resulted from the statistics of the 
$p+^4$He scattering events.
Future development of the NMR system would be required for determining 
the absolute polarization more precisely without losing beam time.

\bibliographystyle{prsty}

\end{document}